\documentclass[twocolumn,showpacs,amsmath,amssymb,prb,floatfix,aps]{revtex4-1}
\usepackage{graphicx}
\usepackage{dcolumn}
\usepackage{bm}
\begin{document}
\title{Electric-field induced penetration of
edge states \\
at the interface between 
monolayer and bilayer graphene 
}
\author{Yasumasa Hasegawa$^1$
and Mahito Kohmoto$^2$}
\affiliation
{$^1$Department of Material Science, Graduate School of Material Science, 
University of Hyogo, \\
3-2-1 Kouto, Kamigori, Hyogo, 678-1297, Japan \\
$^2$Institute for Solid State Physics, University of Tokyo, 
5-1-5 Kashiwanoha, Kashiwa, Chiba 277-8581, Japan
}
%
\date{July 7, 2011, revised \today}

\begin{abstract}
The edge states in the hybrid system of single-layer 
and double-layer graphene are studied 
in the tight-binding model theoretically.
The edge states 
in one side of the interface between single-layer 
and double-layer graphene are shown to penetrate into the
single-layer region when the perpendicular 
electric field is applied, while they are localized
in the double-layer region without electric field.
The edge states in another side of the interface 
are localized in the double-layer
region independent of the electric field. 
This field-induced penetration 
of the edge states can be applied to switching devices.
We also find a new type 
of the edge states at the boundary between single-layer 
and the double-layer graphene.

\end{abstract}

\maketitle

\section{Introduction}
Recently, single-layer 
 and double-layer 
graphene have been studied 
both theoretically and experimentally\cite{RevModPhys.81.109}, 
because of the
interesting properties such as the Dirac points\cite{Novoselov2005,Zhang2005},
anomalous Hall effect\cite{Novoselov2005,Zhang2005,Gusynin2005,Hasegawa2006}, 
and the edge 
states\cite{Fujita1996,Nakada1996,Kohmoto2007,Castro2007,Castro2008PRL,%
Ritter2009,
Tao2011}.
The double-layer graphene has attracted peculiar interest\cite{Castro2010JoP}
due to a band gap controlled 
by the electric field,
which has been 
predicted\cite{McCann2006PRL,McCann2006PRB} 
and observed\cite{Ohta2006,Castro2007,Oostinga2008,Mak2009,%
Kuzmenko2009,Zhang2009,Li2009}.

The edge states in the single-layer graphene
 and double-layer graphene
have been studied by many authors. 
In the single-layer graphene 
the edge states appear at the zigzag edges\cite{Fujita1996,Nakada1996}
and bearded edges. If the system is anisotropic, 
the edge states also exist at the armchair edges\cite{Kohmoto2007}. 
The edge states in the double-layer graphene has been 
studied\cite{Castro2007,Castro2008PRL,Castro2008EPL}. 
The edge states at the interface between single-layer 
and double-layer graphene
have also been studied. Transmission across the boundary has been studied 
theoretically using the effective-mass 
approximation\cite{Nilson2007,Nakanishi2010,Gonzalez2010}, 
edge states have been studied
 theoretically\cite{Castro2008EPL,Hu2011}, 
and quantum oscillations have been observed
in the interface\cite{Puls2009}. Vacancy-induced localized states
in the multilayer graphene has been proposed\cite{Castro2010}.

In this paper we study the edge states in the hybrid system of 
single-layer and double-layer graphene as shown in Fig.~\ref{figlatbising2}.
We focus on the edge states localized in the boundary between
the single-layer and the double-layer regions.
We obtain the new edge states localized in one side of the interface
between single-layer and double-layer regions with energy $E \ne 0$.
We show an interesting property 
that the edge states at the boundary between the single-layer 
and the double-layer regions
penetrate into the single-layer region when the electric field is 
applied perpendicular to the layers.
\section{model}
%
%
\begin{figure}[b]
\begin{center}
\vspace{0.0cm}
\includegraphics[width=0.45\textwidth]{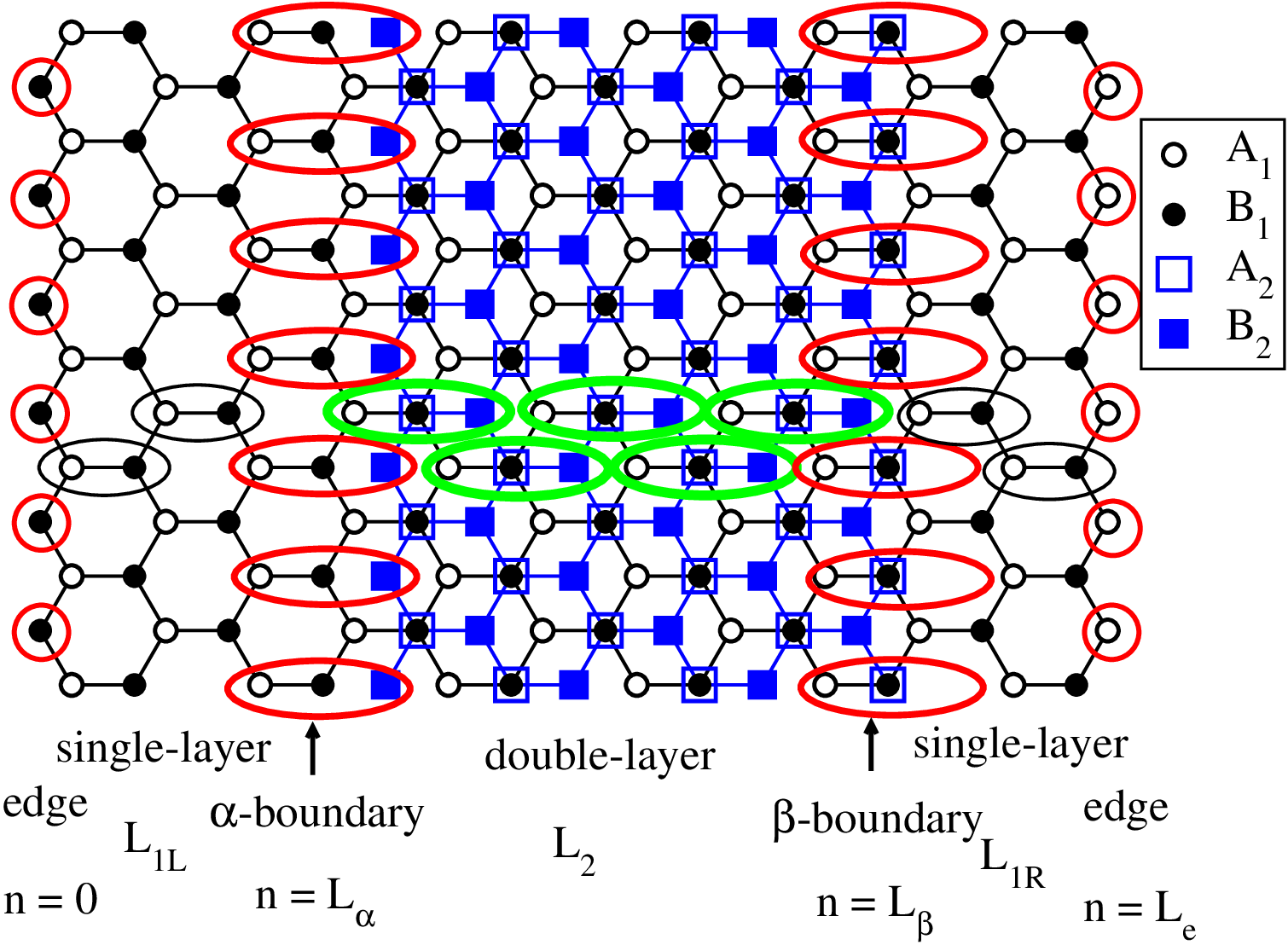}
\end{center}
\caption{(color online).
Single-double-single layer graphene.
Thin black ellipses are the doublets of ($A_1$, $B_1$) 
sites in the single-layer region, and
thick green ellipses are the quartets of 
($A_1$, $B_1$, $A_2$, $B_2$) sites in the double-layer region. 
Thick red ellipses are the triplets of ($A_1$, $B_1$, $B_2$) and 
($A_1$, $B_1$, $A_2$) 
sites (left and right, respectively)
in the boundary between single layer and double layer
of the $\alpha$- and $\beta$-types, respectively.
Thick red circles are the zigzag edges.} 
\label{figlatbising2}
\end{figure}

We assume zigzag edges in both the first and the second layers.
Each layer has two sublattices,
which we call $A_1$, $B_1$, $A_2$ and $B_2$, as shown 
in Fig.~\ref{figlatbising2}.  
The left part and the right part of the single-layer regions have
$L_{1L}$ and $L_{1R}$ pairs of $A_1$ and $B_1$ sublattices 
in the x-direction. 
In the double-layer region there are $L_2$ quartets of
$(A_1, B_1, A_2, B_2)$ sublattices. 
The left edge in the left single-layer region 
has only $B_1$ sublattice, which is labeled as $n=0$,
and the right edge in the right single-layer region 
($A_1$ sublattice) is labeled as $n=L_{e}
\equiv L_{1L}+L_{1R}+L_2+3$.
There are two boundaries between single-layer 
 and the double-layer regions.
These two boundaries are 
different from each other\cite{Castro2008EPL,Gonzalez2010,Nakanishi2010,Hu2011}.
 One of the boundaries has
$A_1, B_1$ and $B_2$ sublattices and the other has
$A_1, B_1$ and $A_2$ sublattices. We call these boundaries as 
$\alpha$-boundary 
and $\beta$-boundary, respectively.
The position of the $\alpha$-boundary is $n=L_{\alpha} \equiv L_{1L}+1$, and
the position of the $\beta$-boundary is $n=L_{\beta}\equiv L_{1L}+L_2+2$.
Note that $\alpha$-boundary and the $\beta$-boundary always appear 
as a pair if two boundaries are parallel. 
We assume that the left boundary is the $\alpha$ type and
the right boundary is the $\beta$ type.

We adopt the tight-binding model, where
the hoppings between the nearest sites
in the layer ($A_1$-$B_1$ and $A_2$-$B_2$) are taken to be 
 $t$ and the interlayer hoppings between the
nearest sites ($B_1$-$A_2$)  are taken to be $t_{\perp}$.
 We take into account the energy difference
between layers ($\epsilon_1$ and $\epsilon_2$), which is controlled 
by the electric field perpendicular to the layers.

We apply the same method which we have 
used in studying the edge states in the single-layer 
graphene\cite{Kohmoto2007}. Imposing the periodic boundary conditions
in the $y$ direction, we can take the wave number $k_y$ 
as the quantum number.
For each $k_y$ the Hamiltonian is written as a 
$L_{sds} \times L_{sds}$ matrix,
where $L_{sds} \equiv 2 L_{1L} + 4 L_2 +2 L_{1R}+8$.
The eigenstates ($\Psi$) in 
the Schr\"odinger equation
 ($H \Psi = E \Psi$)
are  vectors with 
$L_{sds}$ components (the wave functions 
for the first layer ($\Psi_{B_1,0}$,  $\Psi_{A_1,L_e}$,
 $\Psi_{A_1,n_1}$, and
$\Psi_{B_1,n_1}$,  where
$1 \leq n_1 \leq L_e-1$) and the
wave functions for the second layer
($\Psi_{B_2,L_{\alpha}}$, $\Psi_{A_2,L_{\beta}}$,
$\Psi_{A_2,n_2}$, and $\Psi_{B_2,n_2}$,
where $L_{\alpha}+1 \leq n_2 \leq L_{\beta}-1$)).

In the single-layer region ($1 \leq n \leq L_{\alpha}-1$,  or 
$L_{\beta}+1 \leq n \leq L_e-1$), the Schr\"odinger equation
is written as,
\begin{align}
     -2t \cos \frac{k_y}{2} \Psi_{B_1,n-1} - t \Psi_{B_1,n}
  &= (E-\epsilon_{1}) \Psi_{A_1,n}, 
\label{eqsingle1}\\
    -2t \cos \frac{k_y}{2} \Psi_{A_1,n+1} - t \Psi_{A_1,n} 
  &= (E-\epsilon_{1}) \Psi_{B_1,n} .
\label{eqsingle2}
\end{align}
 
The equations in the region of double layer ($L_{\alpha}+1 \leq n \leq
L_{\beta}-1$) are given by
\begin{align}
 -2t \cos \frac{k_y}{2} \Psi_{B_1, n-1} -t  \Psi_{B_1, n} 
&= (E-\epsilon_{1}) \Psi_{A_1,n},
\label{eqdouble1}\\
 -2t  \cos \frac{k_y}{2} \Psi_{A_1, n+1} -t  \Psi_{A_1, n} &
\nonumber \\
- t_{\perp} \Psi_{A_2,n}
&= (E-\epsilon_{1}) \Psi_{B_1,n},
\label{eqdouble2}\\
 -2t  \cos \frac{k_y}{2} \Psi_{B_2, n-1} -t  \Psi_{B_2, n} &
\nonumber \\
 -t_{\perp} \Psi_{B_1,n}  
&= (E-\epsilon_{2}) \Psi_{A_2,n},
\label{eqdouble3}\\
 -2t  \cos \frac{k_y}{2} \Psi_{A_2, n+1} -t  \Psi_{A_2, n} 
&= (E-\epsilon_{2}) \Psi_{B_2,n}.
\label{eqdouble4}
\end{align}

At the left  edge of the single-layer region
we obtain the equation to be Eq.~(\ref{eqsingle2}) 
with $n=0$ and $\Psi_{A_1,0}=0$, since there are no $A_1$ sublattice at the 
left edge in the single-layer region, i.e.,
\begin{equation}
    -2t \cos \frac{k_y}{2} \Psi_{A_1,1} 
  = (E-\epsilon_{1}) \Psi_{B_1,0} .
\end{equation}
Similarly, we obtain the equation 
at the right edge of the single-layer region
to be Eq.~(\ref{eqsingle1}) with $n=L_e$ and $\Psi_{B_1,L_e}=0$,
\begin{equation}     -2t \cos \frac{k_y}{2} \Psi_{B_1,L_e-1} 
  = (E-\epsilon_{1}) \Psi_{A_1,L_e}, 
\end{equation}

At the $\alpha$-boundary, the equations are
obtained by taking $n=L_{\alpha}$ and $\Psi_{A_2,L_{\alpha}}=0$
in Eqs.~(\ref{eqdouble1}), (\ref{eqdouble2}), and (\ref{eqdouble4}),
since there are no $A_2$ sublattices at the $\alpha$-boundary.
\begin{align}
 -2t \cos \frac{k_y}{2} \Psi_{B_1, L_{\alpha}-1} -t  \Psi_{B_1, L_{\alpha}} 
&= (E-\epsilon_{1}) \Psi_{A_1,L_{\alpha}},
\label{eqalpha1}\\
 -2t  \cos \frac{k_y}{2} \Psi_{A_1, L_{\alpha}+1} -t  \Psi_{A_1, L_{\alpha}} 
&= (E-\epsilon_{1}) \Psi_{B_1,L_{\alpha}},
\label{eqalpha2}\\
 -2t  \cos \frac{k_y}{2} \Psi_{A_2, L_{\alpha}+1} 
&= (E-\epsilon_{2}) \Psi_{B_2,L_{\alpha}}.
\label{eqalpha4}
\end{align}

The equations at the $\beta$-boundary  are
obtained by taking $n=L_{\beta}$ and $\Psi_{B_2,L_{\beta}}=0$, 
in Eqs.~(\ref{eqdouble1}), (\ref{eqdouble2}), and (\ref{eqdouble3}). 
Explicitly, the equations at the $\beta$-boundary are given by
\begin{align}
 -2t \cos \frac{k_y}{2} \Psi_{B_1, L_{\beta}-1} -t  \Psi_{B_1, L_{\beta}} 
=&  (E-\epsilon_{1}) \Psi_{A_1,L_{\beta}},
\label{eqbeta1}\\
 -2t  \cos \frac{k_y}{2} \Psi_{A_1, L_{\beta}+1} -t  \Psi_{A_1, L_{\beta}}
 &\nonumber \\ 
-t_{\perp} \Psi_{A_2, L_{\beta}}
=& (E-\epsilon_{1}) \Psi_{B_1,L_{\beta}},
\label{eqbeta2}\\
 -2t  \cos \frac{k_y}{2} \Psi_{B_2, L_{\beta}-1} 
 -t_{\perp} \Psi_{B_1,L_{\beta}}  
=&  (E-\epsilon_{2}) \Psi_{A_2,L_{\beta}}.
\label{eqbeta3}
\end{align}

By taking $E=\epsilon_1$, we obtain that Eqs.~(\ref{eqsingle1}) and 
(\ref{eqsingle2}) are two independent equations 
for $\Psi_{B_{1},n}$ and $\Psi_{A_{1},n}$, respectively, in the
single-layer regions.
When $\epsilon_{1}=\epsilon_{2}=0$,
Eqs.~(\ref{eqdouble1}) - (\ref{eqdouble4}) become 
two sets of coupled equations for ($\Psi_{A_1,n}$, $\Psi_{A_2,n}$)
and ($\Psi_{B_1,n}$, $\Psi_{B_2,n}$) in the double-layer
region by taking $E=0$. 
However, if $\epsilon_1 \neq \epsilon_2$,
these equations cannot be separated into the independent equations
for any $E$.
This is the origin of the field-induced penetration of the edge states 
into the first-layer region
at the $\beta$-boundary, as we will show below.

\section{strictly localized states at $k_y=\pi$}

Since $\cos k_y/2 =0$ at $k_y=\pi$, 
Eqs.~(\ref{eqsingle1}) - (\ref{eqbeta3})
are the equations within the same group of $n$, i.e. the states at $k_y=\pi$
are strictly localized 
at the ellipses or circles in Fig.~\ref{figlatbising2},
as in the single-layer graphene\cite{Kohmoto2007}.

The energies at $k_y=\pi$ in the single-layer regions are obtained 
from Eqs~(\ref{eqsingle1}) and (\ref{eqsingle2}),
as
\begin{equation}
E_{s,\pm} =  \pm t + \epsilon_1, 
\end{equation}
with $(L_{1L}+L_{1R})$-fold degeneracy. 
The energies  of the strictly localized states
 in the double-layer region are obtained as
the eigenvalues of the matrix
\begin{equation}
 M_d = \left( \begin{array}{cccc}
 \epsilon_1 & -t         & 0          & 0         \\
 -t         & \epsilon_1 & -t_{\perp} & 0         \\
 0          & -t_{\perp} & \epsilon_2 & -t        \\
 0          &  0         & -t         & \epsilon_2
 \end{array} \right),
\end{equation}
and they are obtained to be
\begin{align}
& E_{d,\pm,\pm} =   \frac{\epsilon_1+\epsilon_2}{2} \nonumber \\
&\pm \sqrt{(\frac{\Delta \epsilon}{2})^2+ t^2+ \frac{t_{\perp}^2}{2} 
\pm \sqrt{(\Delta \epsilon)^2 t^2 + t^2 t_{\perp}^2  +\frac{t_\perp^4}{4}}},
\end{align}
where $\Delta \epsilon =\epsilon_1-\epsilon_2$,
with $L_{2}$-fold degeneracy.

At the left and the right edges we obtain the energy as 
\begin{equation}
E_{L} = E_{R} =\epsilon_1.
\end{equation}
At the $\alpha$-boundary we obtain 
the energies of the strictly localized states
as the eigenvalues of the matrix
\begin{equation}
 M_{\alpha} = \left( \begin{array}{ccc}
 \epsilon_1 & -t         & 0         \\
 -t         & \epsilon_1 & 0         \\
 0          &  0         & \epsilon_2
 \end{array} \right),
\end{equation}
which are obtained as
\begin{align}
E_{\alpha,0} = \epsilon_2 , 
\end{align}
and
\begin{align}
E_{\alpha,\pm} =  \pm t + \epsilon_1.
\end{align}
The energies of the strictly localized states 
at the $\beta$-boundary  
are obtained as the 
eigenvalues of the matrix,
\begin{equation}
M_{\beta}=
 \left( \begin{array}{ccc}
  \epsilon_1 & -t & 0\\
  -t & \epsilon_1 & - t_{\perp} \\
  0  & - t_{\perp} & \epsilon_2
\end{array} \right).
\end{equation}
When $|\epsilon_1| \ll t$ and  $|\epsilon_2| \ll t$, 
we obtain the energies as
\begin{align}
  E_{\beta 0} 
 =& \epsilon_2
 + \frac{t_{\perp}^2 }{t^2+t_{\perp}^2} \Delta \epsilon\nonumber \\
 &+ \frac{t^4 t_{\perp}^2}{(t^2+t_{\perp}^2)^4} (\Delta \epsilon)^3 
+O((\Delta \epsilon)^5)
 \\  
  E_{\beta \pm} 
 =& \pm \sqrt{t^2+t_{\perp}^2} +\epsilon_2
+ \frac{(2 t^2+t_{\perp}^2)}{2(t^2+t_{\perp}^2)}\Delta \epsilon
\nonumber \\
 &\pm \frac{t_{\perp}^2 (4 t^2+t_{\perp}^2)}{8 (t^2+t_{\perp}^2)^{5/2}}
(\Delta \epsilon)^2 +O((\Delta \epsilon)^3).
\end{align}
The eigenstates with the eigenvalues $E_{\beta,0}$ and $E_{\beta,\pm}$ are
 obtained as
\begin{equation}
 \left( \begin{array}{c}
 \Psi_{A_1,L_\beta} \\
 \Psi_{B_1,L_\beta} \\
 \Psi_{A_2,L_\beta} 
 \end{array}\right) =
 \Psi_{A_2,L_\beta} 
 \left( \begin{array}{c}
 -\frac{t_\perp}{t} +O((\Delta \epsilon)^2))\\
 -\frac{t_\perp \Delta \epsilon}{t^2+t_{\perp}^2}
 +O((\Delta \epsilon)^3))\\
 1 
 \end{array}\right),
\label{eqbeta00}
\end{equation}
and
\begin{align}
& \left( \begin{array}{c}
 \Psi_{A_1,L_\beta} \\
 \Psi_{B_1,L_\beta} \\
 \Psi_{A_2,L_\beta} 
 \end{array}\right) 
\nonumber \\
=&
  \Psi_{A_1,L_\beta}
 \left( \begin{array}{c}
 1 \\
 \mp \frac{\sqrt{t^2+t_{\perp}^2}}{t}+
 \frac{t_{\perp}^2 \Delta \epsilon}{2t(t^2+t_{\perp}^2)} 
+O((\Delta \epsilon)^2))\\
 \frac{t_{\perp}}{t} 
   \mp \frac{t_\perp \Delta \epsilon}{t\sqrt{t^2+t_{\perp}^2}}
+O((\Delta \epsilon)^2))
 \end{array}\right),
\label{eqbcbetapm}
\end{align}
respectively.

Since $E_L$, $E_R$, $E_{\alpha,0}$, $E_{\beta,0}$, 
and $E_{\beta,\pm}$ are different from the
energies of the macroscopically degenerate states 
($E_{s,\pm}$ and $E_{d,\pm,\pm}$), the eigenstates with these energies
become the well-defined edge states at $k_y \approx \pi$, 
as we will show below. 

\section{edge states without perpendicular electric field}
\subsection{edge states with $E=0$}
%
\begin{figure}[bt]
\flushleft{ (a) \hfill (b) \hfill \ } \vspace{-0.0cm}\\
\begin{center}
\includegraphics[width=0.23\textwidth]{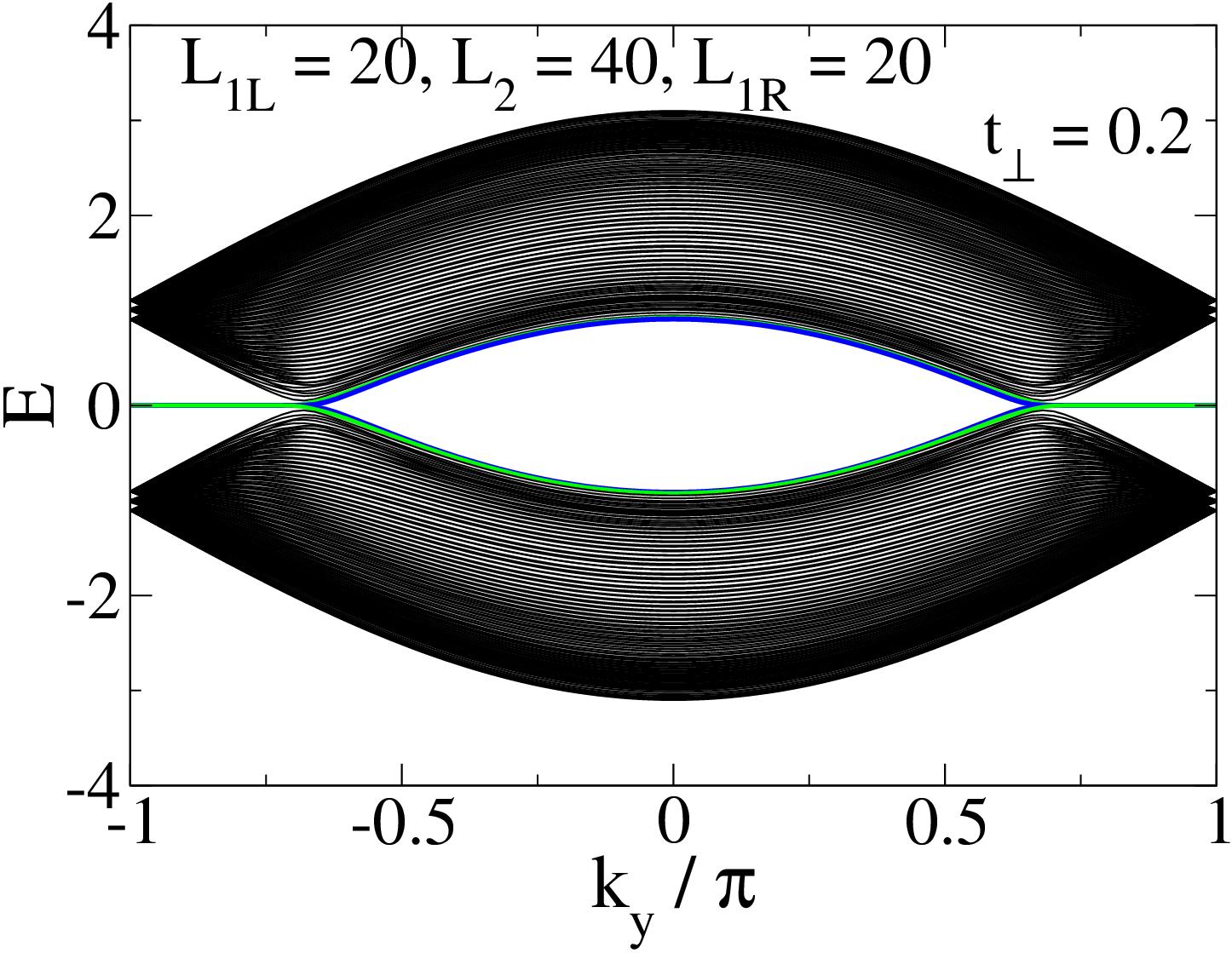} \hfill
\includegraphics[width=0.23\textwidth]{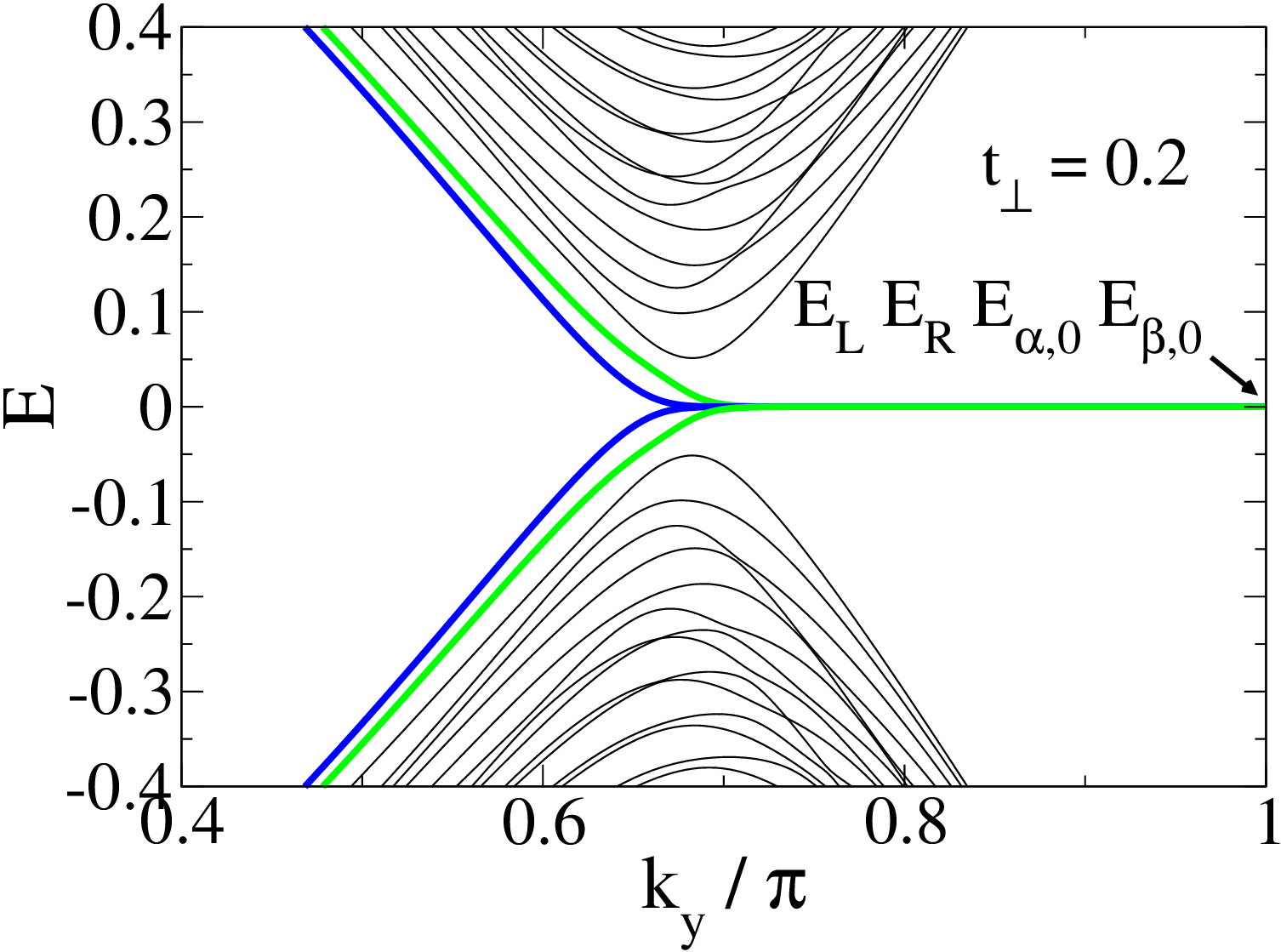}
\end{center}
\flushleft{ (c) \hfill (d) \hfill \ } \vspace{-0.1cm} \\
\begin{center}
\includegraphics[width=0.23\textwidth]{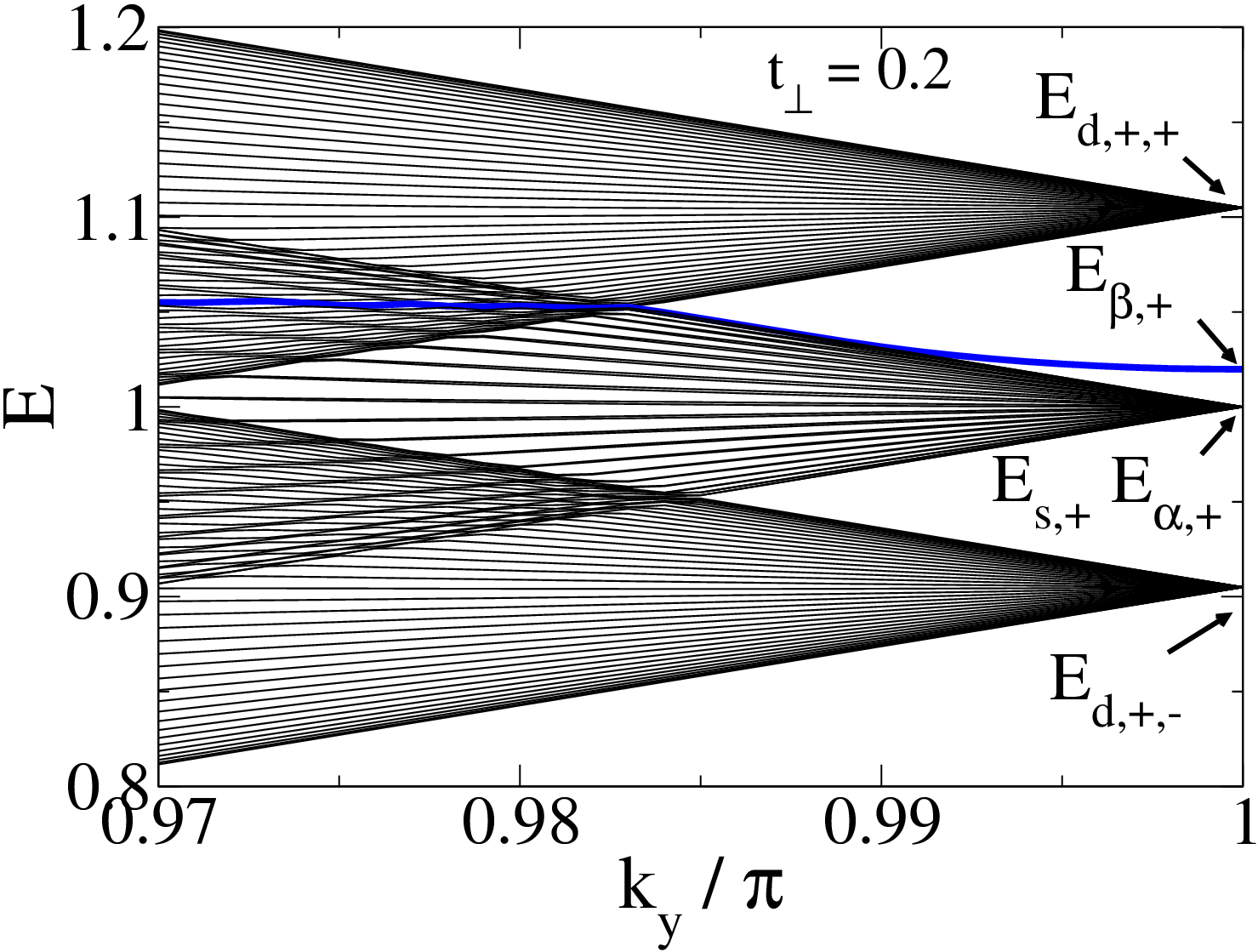} \hfill
\includegraphics[width=0.23\textwidth]{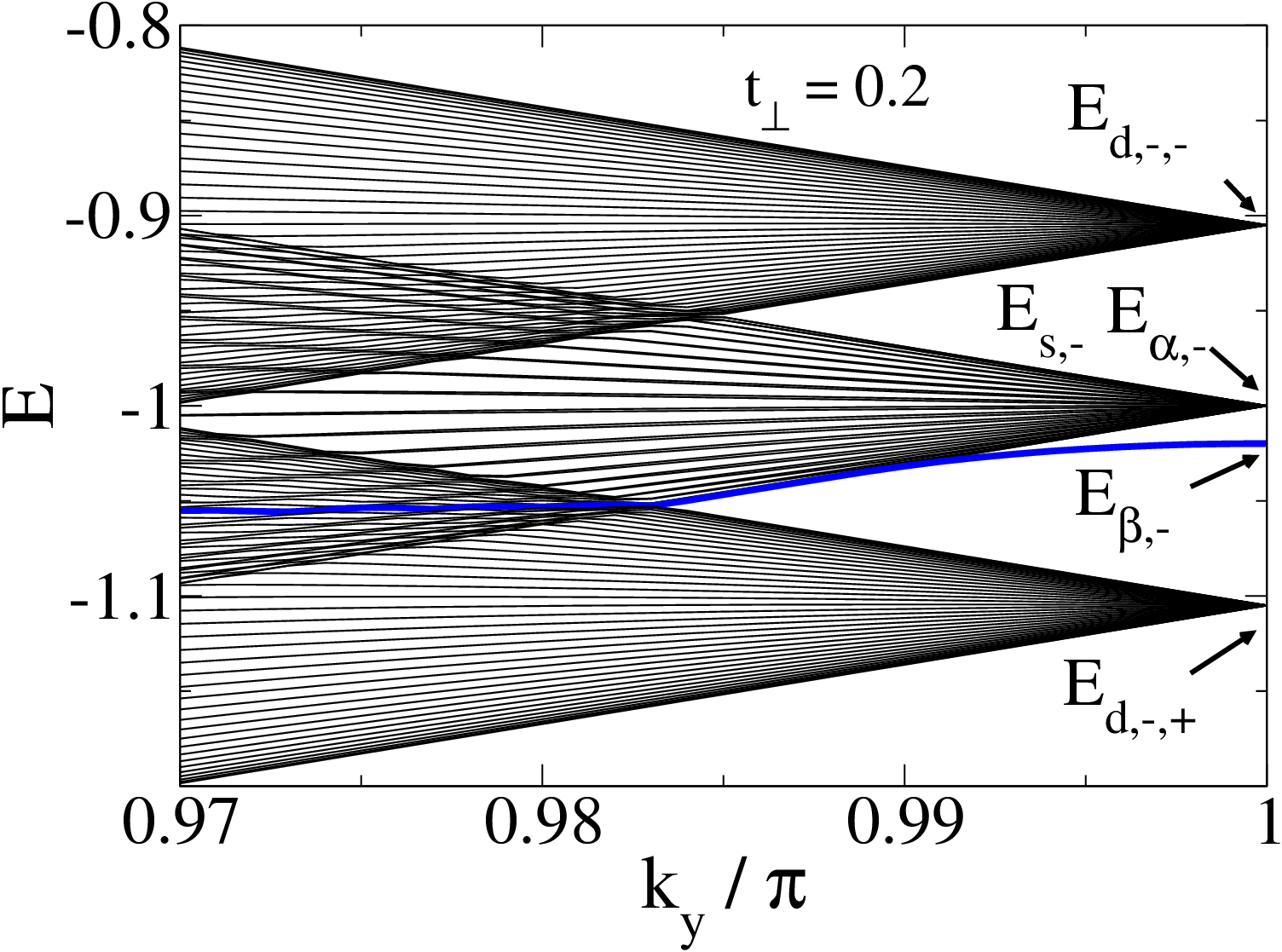}\\
\end{center}
\caption{(color online).
Energy as a function of $k_y$ for single-double-single graphene.
There are four edge states at $E=0$
as shown in (b), two of them are 
edge states at the left and the 
right zigzag edges of single-layer regions
(see Fig.~\ref{figwavesdse000} (a)). 
The other two edge states at $E=0$ are
the edge states in the double-layer region 
at the $\alpha$ and the $\beta$ boundaries
(see Fig.~\ref{figwavesdse000} (b) and (c)).
 There exist other two edge states at 
$E=E_{\beta,+}$ and $E=E_{\beta,-}$
near $|k_y| =\pi$ as shown in (c) and (d). 
}
\label{figsds}
\end{figure}
%
%
\begin{figure}[bt]
\flushleft{(a)\hfill (b) \hfill \ }  \\  \vspace{-0.2cm}
\begin{center}
\includegraphics[width=0.23\textwidth]{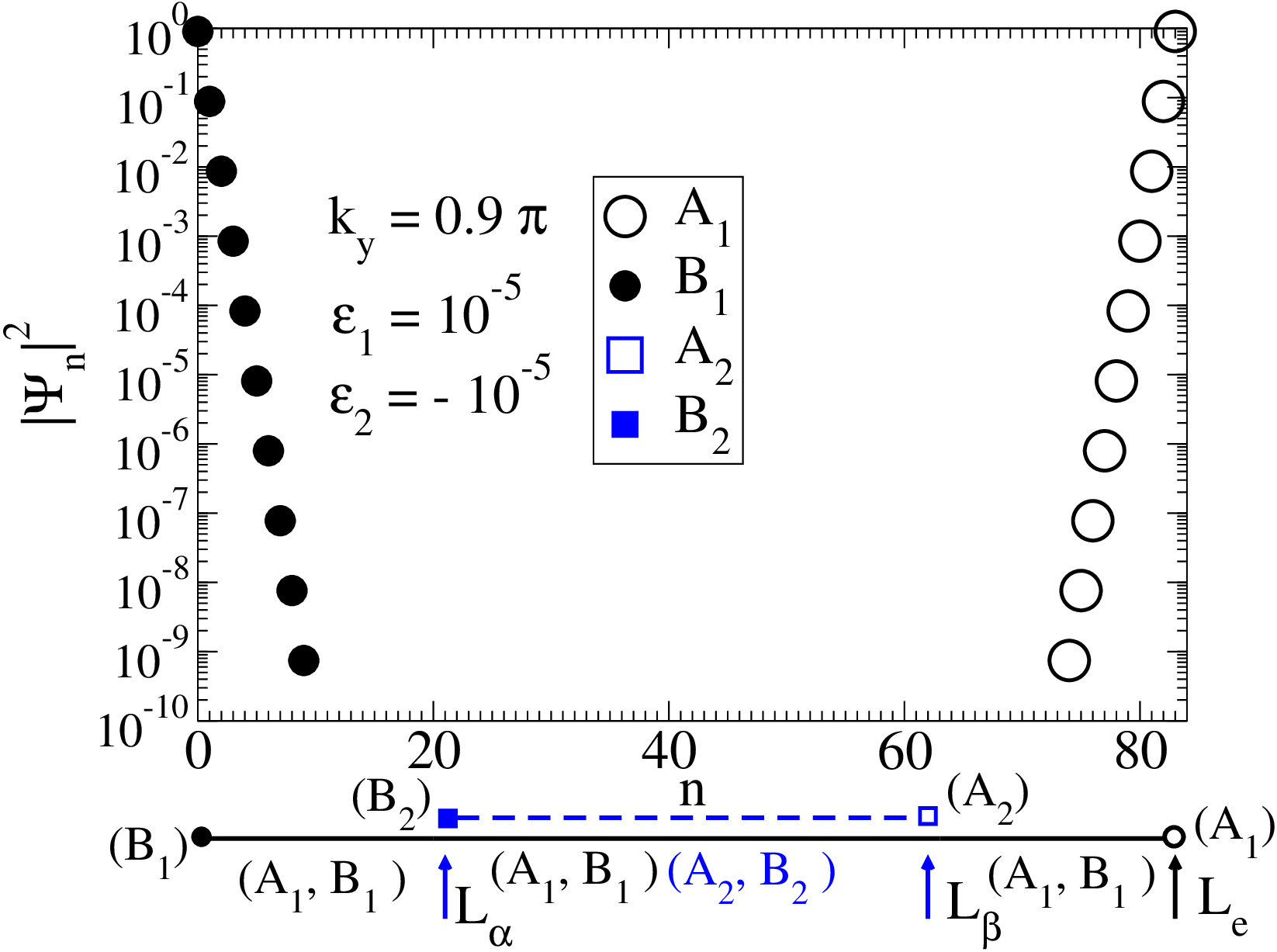} \hfill
\includegraphics[width=0.23\textwidth]{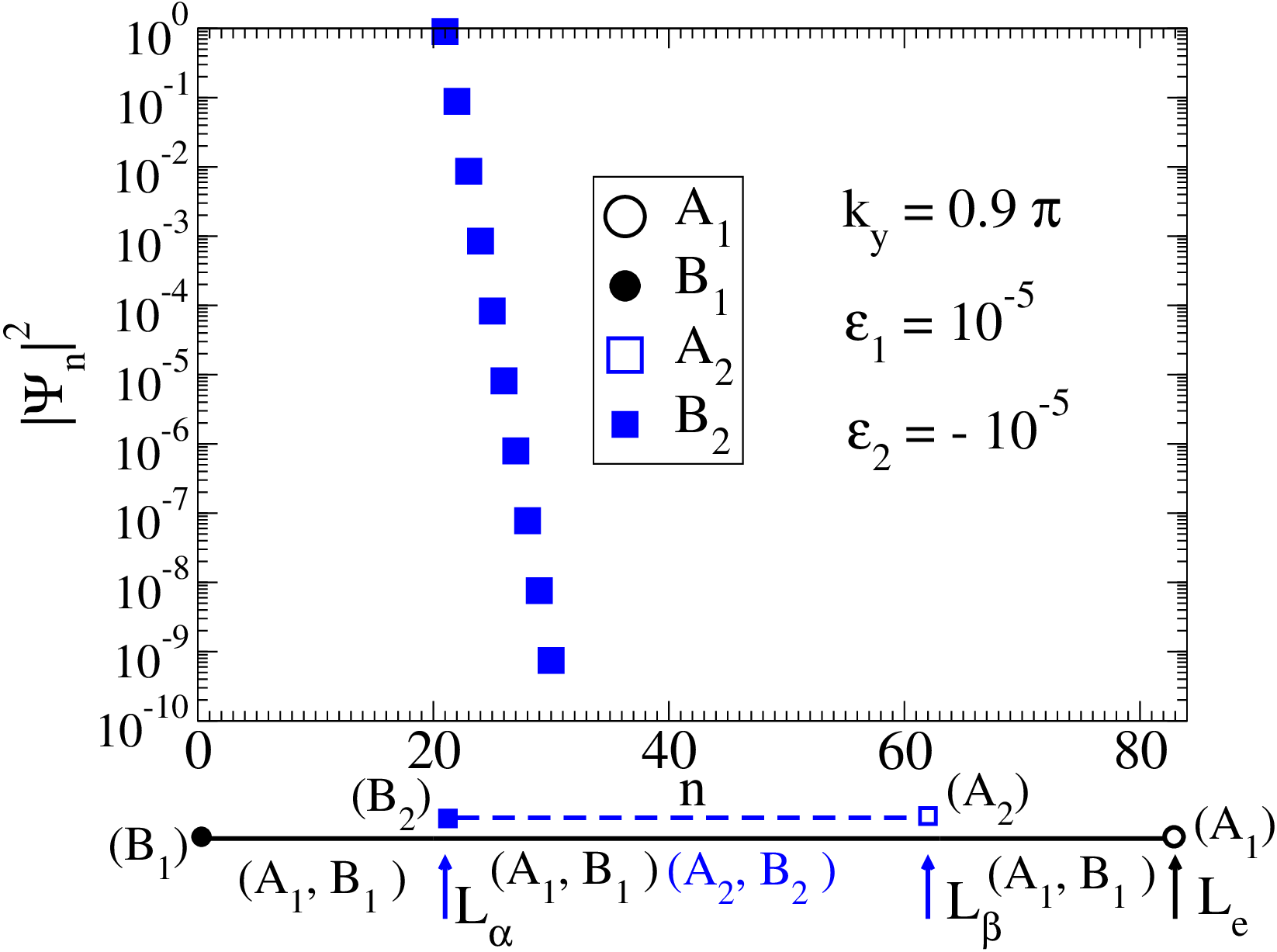} 
\end{center}
\flushleft{(c) \hfill (d) \hfill \ }  \\ \vspace{-0.2cm}
\begin{center}
\includegraphics[width=0.23\textwidth]{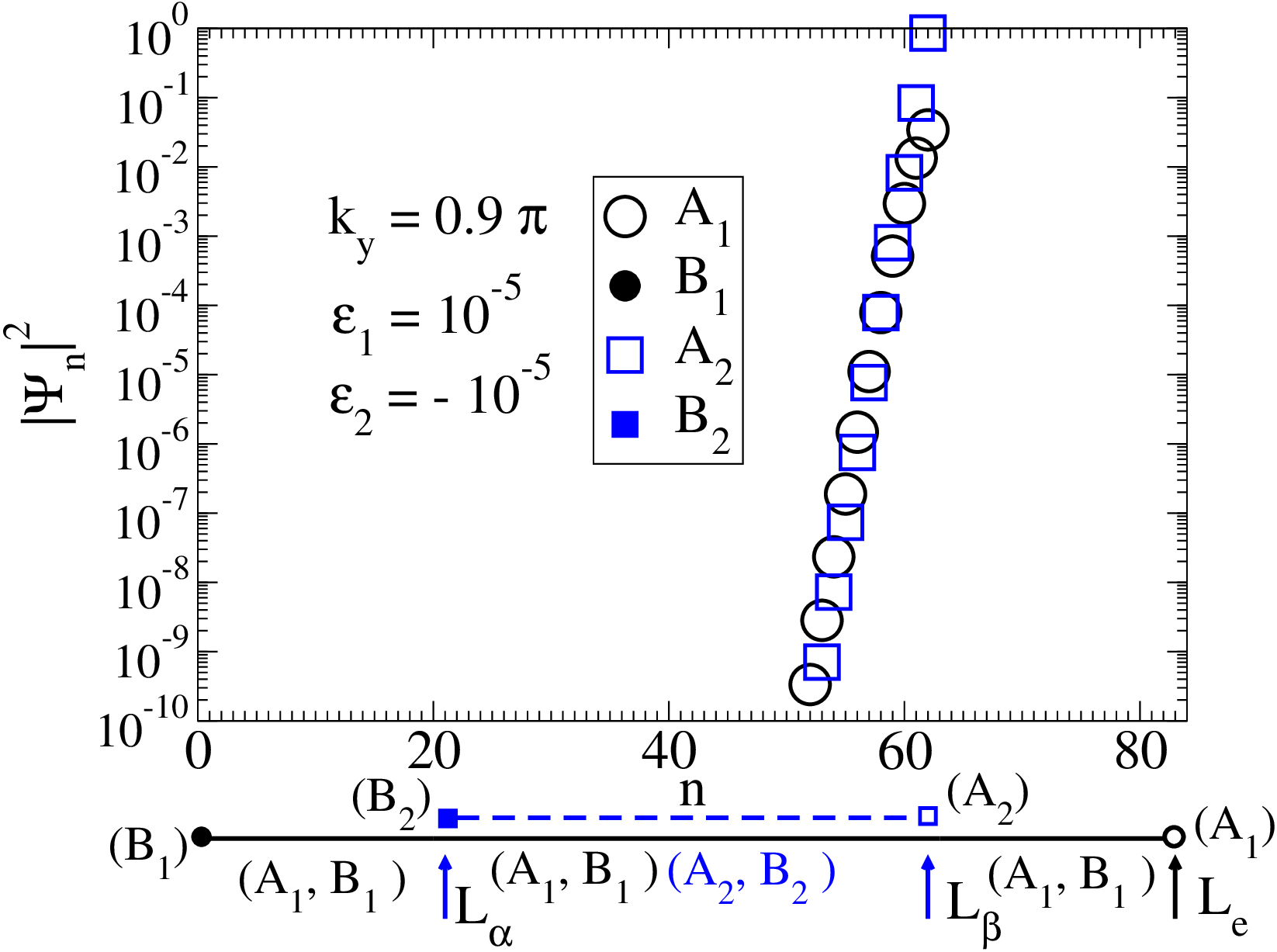} \hfill
\includegraphics[width=0.23\textwidth]{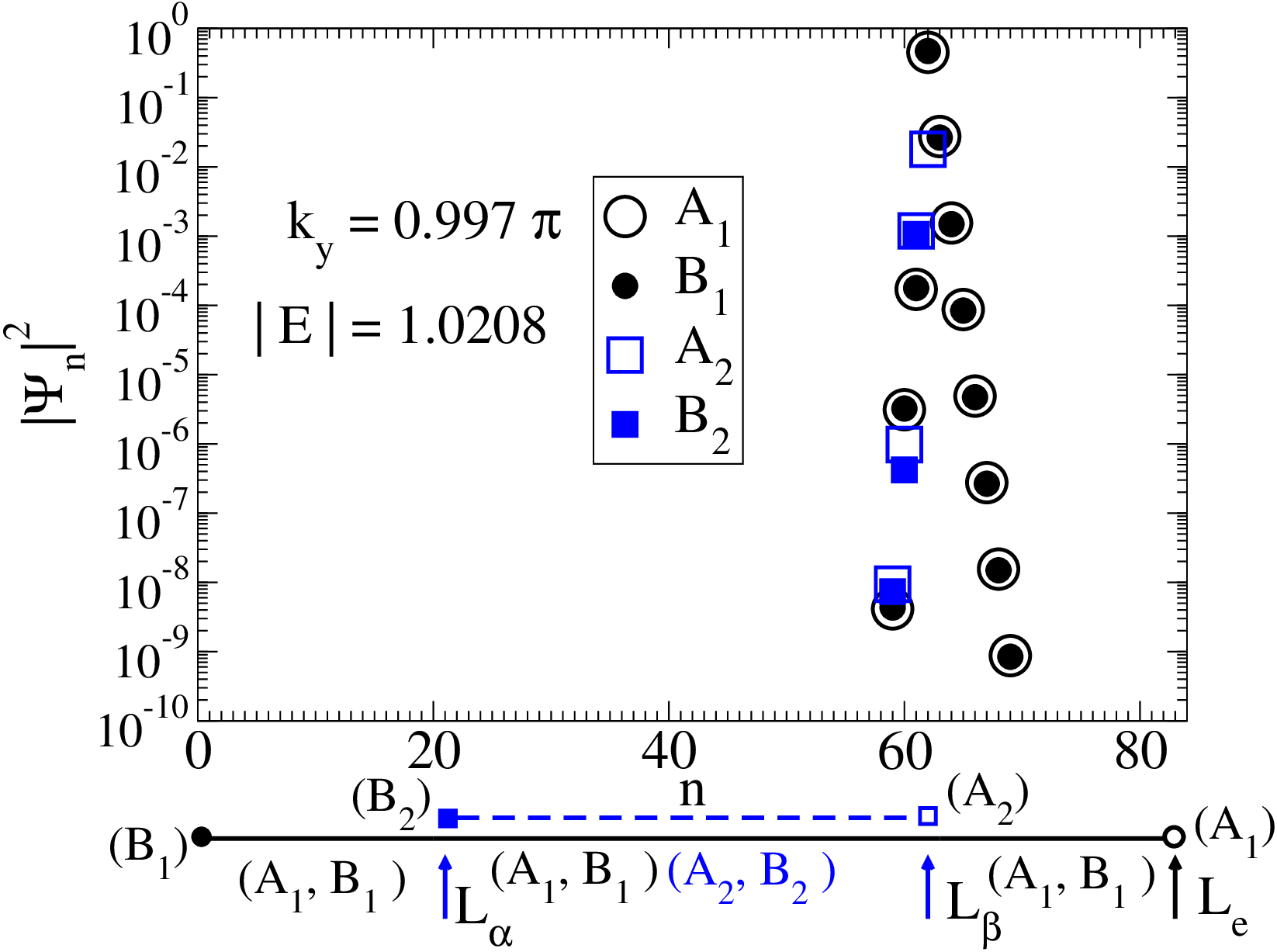}
\end{center}
\caption{(color online).
Edge states localized (a) at the left and the right edges,
(b) at the $\alpha$-boundary and (c) at the $\beta$-boundary  
 at $E=0$ and $k_y = 0.9 \pi$. (d) is the edge state localized 
at the $\beta$-boundary at $E \approx E_{\beta,+}$
or $E \approx E_{\beta,-}$ and 
$k_y = 0.997 \pi$.}
\label{figwavesdse000}
\end{figure}
First, we study the edge states in the case of
no external electric field ($\epsilon_1=\epsilon_2=0$).
We plot the energy as a function of $k_y$ in Fig.~\ref{figsds},
where we
take $t=1$, $t_{\perp}=0.2$, $L_{1L}=20$, $L_{1R}=20$, $L_2=40$, and 
$\epsilon_{1}=\epsilon_{2}=0$.
As shown in the previous section, 
there are four states which have $E=0$ when $\epsilon_1=\epsilon_2=0$ and
$k_y=\pi$, 
i.e., $E_L$, $E_R$, $E_{\alpha,0}$, and $E_{\beta,0}$. 
Two of them are edge states localized at each edge 
in the single-layer regions ($n=0$ and $n=L_e$) for $|k_y| > 2 \pi/3$,
same as the single-layer graphene\cite{Kohmoto2007}.
The edge state localized at the left edge of the single layer is given by 
\begin{equation}
  \Psi_{B_1,n}  =
\left(-2 \cos \frac{k_y}{2} \right)^n \Psi_{B_1,0},
\label{eqb1sj}
\end{equation}
where $0 \leq n \leq L_{1L}$ and  other components of $\Psi$ are zero.
The edge state localized at the right edge of the single layer is given by
\begin{align}
  \Psi_{A_1,L_e-j} &=\left(-2 \cos \frac{k_y}{2} \right)^j \Psi_{A_1,L_e},
\end{align}
where $0 \leq j \leq L_{1R}$ and other 
components of $\Psi$ are zero.

The other edge states with $E=0$ are localized at the
$\alpha$ and $\beta$ boundaries. As shown in Appendix \ref{appA},
the edge state localized at the $\alpha$-boundary are obtained as
\begin{equation}
 \left( \begin{array}{c}
  \Psi_{B_1,L_\alpha+j} \\
  \Psi_{B_2,L_\alpha+j} 
\end{array} \right) = 
\left( \begin{array}{c} 
0 \\
\left( -2 \cos \frac{k_y}{2} \right)^j
 \Psi_{B_2,L_\alpha}
\end{array} \right),
\end{equation}
where $0 \leq j \leq L_2$ and other components of $\Psi$ are zero.
The edge states at the $\beta$ boundary are obtained as
\begin{equation} 
\left( \begin{array}{c}
 \Psi_{A_1,L_{\beta}-j} \\ \Psi_{A_2,L_{\beta}-j} 
\end{array} \right)
= \left( -2 \cos \frac{k_y}{2} \right)^j \Psi_{A_2,L_\beta}
 \left( \begin{array}{c}
 -\frac{t_{\perp}}{t} (1+j) \\ 1 
\end{array} \right),
\label{eqedgebeta}
\end{equation}
where $0 \leq j \leq L_2$ and other components of $\Psi$ are zero.

These results are consistent with the results obtained 
in the bilayer edge\cite{Castro2008PRL}
 and the graphite steps\cite{Castro2008EPL}.
The edge state at the $\beta$-boundary has the finite 
amplitudes of the wave functions at $A_1$ and $A_2$ sites, while
that at the $\alpha$-boundary has the finite amplitude only at the $B_2$
sites.

We plot the square of the absolute value of
the wave functions 
in Figs.~\ref{figwavesdse000} (a), (b) and (c),
in which we have taken 
$\epsilon_1 =10^{-5}$ and
$\epsilon_2 =-10^{-5}$ in order to lift the
degeneracy of the edge states.
We plot two edge states
together in Fig.~\ref{figwavesdse000} (a), which are localized in the 
left and the right edges of the single-layer regions.
There exist two localized states at each boundary between
single-layer and double-layer, as shown 
in  Fig.~\ref{figwavesdse000} (b) and (c).

\subsection{edge states with $E \neq 0$}
At the $\beta$-boundary
there exist other edge states, which have energy $E \approx E_{\beta,\pm}$
at $|k_y| \approx \pi$.
In Fig.~\ref{figwavesdse000} (d) we plot the amplitudes of 
the wave functions of the edge states 
at $k_y=0.997\pi$ and $E \approx E_{\beta,+}$.
Note that if $\{ \Psi_{A_1,n_1}, \Psi_{B_1,n_1}, \Psi_{A_2,n_2}, 
\Psi_{B_2,n_2} \}$ is the eigenstate with energy $E$, 
$\{ \Psi_{A_1,n_1}, -\Psi_{B_1,n_1}, \Psi_{A_2,n_2}, 
-\Psi_{B_2,n_2} \}$ is also the eigenstate with energy $-E$,
when $\epsilon_1=\epsilon_2=0$.
 Therefore, $|\Psi_{A_1,n}|^2$, 
$|\Psi_{B_1,n}|^2$ $|\Psi_{A_2,n}|^2$, and
$|\Psi_{B_2,n}|^2$ for the edge states with  
$E \approx E_{\beta,-} = - E_{\beta,+}$ are the same as these 
with $E \approx E_{\beta,+}$.

As seen in Fig.~\ref{figsds} (c) and (d), the edge states at $E \approx
E_{\beta, \pm}$ exist at $|k_y| \approx \pi$. 
Although the bulk extended states with the same energy
 $E \approx E_{\beta, \pm}$ also exist at other values of $k_y$,
the density of states has a peak at that energy due to the edge states.
Therefore, the edge states can be
observed as a peak in the differential conductance ($dI/dV$)
 by the spatially resolving 
scanning tunneling spectroscopy (STS)\cite{Ritter2009,Tao2011}
 at the $\beta$-boundary.

These edge states at $E \approx E_{\beta,\pm}$ can be 
understood by considering the equations in the single-layer regions
Eqs.~(\ref{eqsingle1}) and (\ref{eqsingle2})
(see Appendix \ref{appB}).

The existence of these edge states has 
not been known before, as far as we know. 
Although the existence of the edge states at $E \neq 0$ 
has been suggested as a perfectly reflecting states
by Nakanishi et al.\cite{Nakanishi2010},
the pure edge states are
obtained only at $E=0$ in their paper, since they adopted 
the effective-mass scheme, which can be
used only near the Dirac points.

\section{edge states in the
presence of perpendicular electric field}
\begin{figure}[bt]
\flushleft{(a) \hfill (b) \hfill \ }  \\   \vspace{-0.1cm}
\begin{center}
\includegraphics[width=0.23\textwidth]{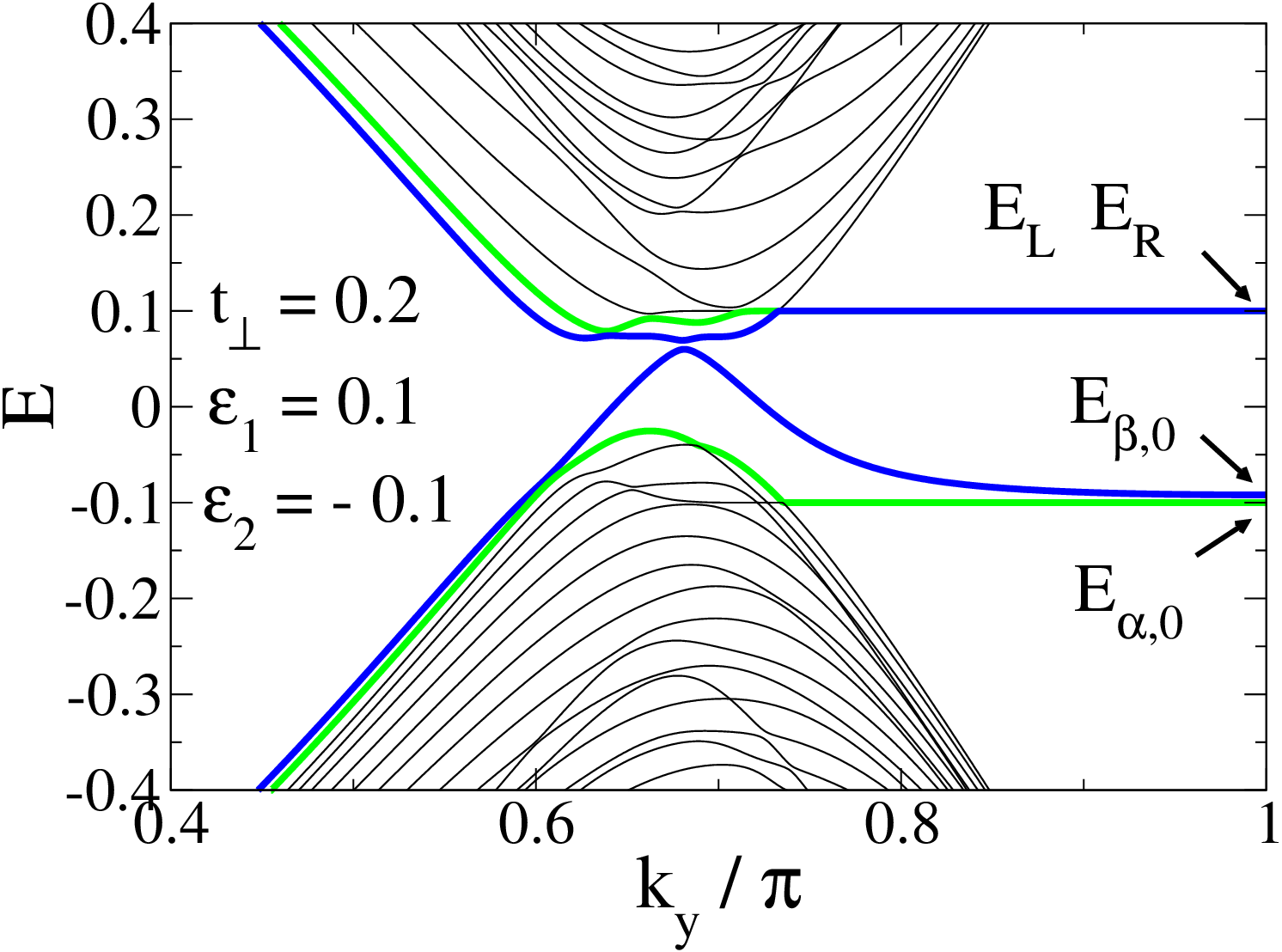} \hfill
\includegraphics[width=0.23\textwidth]{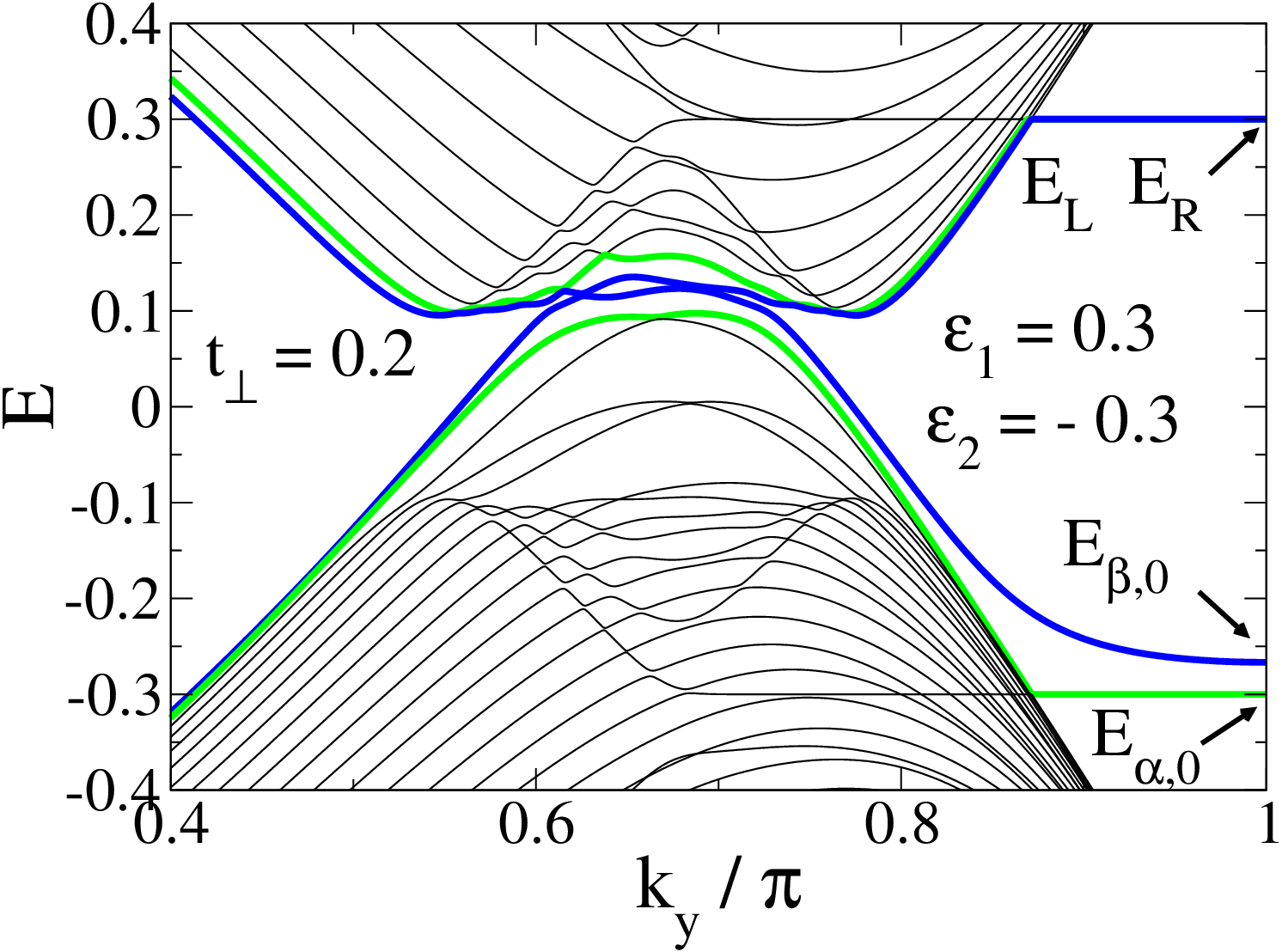}  
\end{center}
\caption{(color online).
Energy of single-double-single layer graphene
as a function of $k_y$ 
with different site energy 
in each layers ((a):$\epsilon_1=-\epsilon_2=0.1 < t_{\perp}$ and 
(b):$\epsilon_1=-\epsilon_2=0.3 > t_{\perp}$.).
}
\label{figsdse01}
\end{figure}

In this section we study the edge states 
in the presence of perpendicular electric field.
When the electric 
field is applied perpendicular to the layers, 
the potential difference between the first and the second 
layers, $(\epsilon_1-\epsilon_2)$, becomes finite.
Even in that case we have the edge states at the left and right edges
in the single-layer regions,
which are given by $E=\epsilon_1$, $\Psi_{A_2,n}=\Psi_{B_2,n}=0$
and $\Psi_{A_1,n}=0$ (the edge states at the left edge)
or $\Psi_{B_1,n}=0$ (the edge states at the right edge).
The edge states at the $\alpha$-boundary, which are given by
$E=\epsilon_2$ and $\Psi_{A_1,n}=\Psi_{B_1,n}=\Psi_{A_2,n}=0$
are also not affected by the perpendicular electric field,
since the edge states at the $\alpha$-boundary is localized only 
in the second layer.
As seen in Fig.~\ref{figsdse01} (a) and (b), the 
states with $E=E_L=E_R=\epsilon_1$
and $E=E_{\alpha,0}=\epsilon_2$ exist 
for $|k_y| \gtrsim 2 \pi/3$ even when they are in the upper 
or lower band.

The edge states at the $\beta$-boundary, however,
 is changed drastically 
by the perpendicular electric field and
they are quite different from
the edge states in the bilayer graphene\cite{Castro2008PRL}.

We study the edge states at the $\beta$-boundary 
with the energy $E \approx E_{\beta,0}$ at $k_y \approx \pi$ in the 
perpendicular electric field in the similar method
given in Appendix \ref{appB}.

We study the single-layer region near the $\beta$-boundary. 
We assume that the energy difference between the first and second layers,
$\epsilon_1-\epsilon_2$ is smaller than the hopping energy in the plane, $t$.
Then  we obtain
\begin{equation}
 |e_0| =\left| \frac{E-\epsilon_1}{t} \right| 
\approx \left| \frac{t (\epsilon_2-\epsilon_1)}{t^2+t_{\perp}^2} \right| <1,
\end{equation}
when $E \approx E_{\beta,0}$.
The  eigenvalues ($\lambda_{\pm}$)
 of the matrix $T$ (Eq.~(\ref{eqmatrixT}))
in the right single layer-region ($L_{\beta}+1  \leq n \leq L_e -1$)
 are real when
\begin{equation}
 |a| < |1-|e_0| | = 1-|e_0|,
\end{equation}
where $a=-\cos  (k_y/2)$  
as discussed in Appendix \ref{appB}.
We expand $\lambda_{\pm}$ and $V$
(defined in Eqs.~(\ref{eqlambda}) and (\ref{eqV})) in $a$, and we obtain
\begin{equation}
 \lambda_+ \approx \frac{1-e_0^2}{a} >1,
\end{equation}
\begin{equation}
 \lambda_- \approx \frac{a}{1-e_0^2} <1,
\end{equation}
and
\begin{equation}
 V \approx \left( \begin{array}{cc}
 2     & -2e_0\\
 -2e_0 & 2
\end{array} \right).
\end{equation}
Therefore, the eigenvector of $T$ with the eigenvalue $\lambda_-$
is the edge state localized 
at the $\beta$-boundary given by
\begin{align}
 \left( \begin{array}{c}
 \Psi_{A_1, L_{\beta}+1+j} \\ \Psi_{B_1,L_{\beta}+1+j} 
 \end{array} \right) 
&= \lambda_{-}^j \Psi_{B_1,L_\beta+1} 
\left( \begin{array}{c}
 -\frac{2e_0}{\sqrt{D}-a^2+e_0^2+1} \\ 1
 \end{array} \right) \nonumber \\
 & \approx
\lambda_{-}^j \Psi_{B_1,L_\beta+1} 
\left( \begin{array}{c}
 -e_0 \\ 1
 \end{array} \right),
\label{eqLbeta1j}
\end{align}
where $0 \leq j \leq L_{1R}$.

If $\Delta \epsilon =0$, we obtain  $\Psi_{B_1,L_\beta}=0$
from Eq.~(\ref{eqbeta00}) and we obtain 
$\Psi_{A_1,L_{\beta}+1}=\Psi_{B_1,L_{\beta}+1}=0$
from Eq.~(\ref{eqLbeta1}). 
In this case, Eq.~(\ref{eqLbeta1j})
shows that the edge states with $E =
E_{\beta,0}=0$ and $|2\cos (k_y/2)| <1$ is localized only 
in the double-layer
region at the $\beta$-boundary, if $\Delta \epsilon =0$,
 as shown Fig.~\ref{figwavesdse000}(c).

If $\Delta \epsilon \neq 0$ due to the perpendicular electric field,
$\Psi_{B_1,L_{\beta}}$ and $\Psi_{B_1,L_{\beta}+1}$ become finite
for the edge states at the $\beta$-boundary, 
resulting in the penetration of the edge 
states into the single-layer region.

%
%
%
%
\begin{figure}[bt]
\flushleft{(a) \hfill (b) \hfill \ }  \\ \vspace{-0.2cm}
\begin{center}
\includegraphics[width=0.23\textwidth]{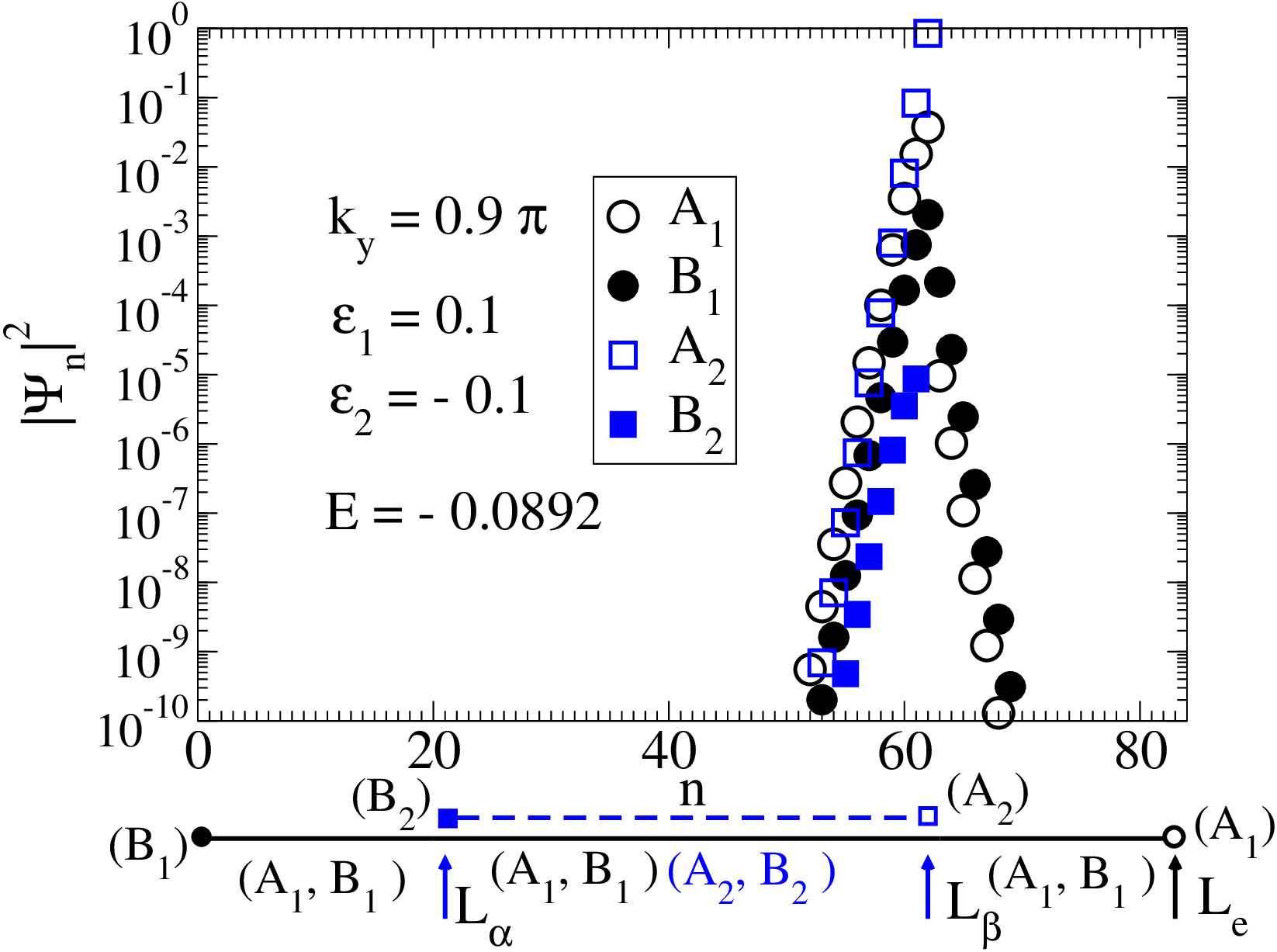} \hfill 
\includegraphics[width=0.23\textwidth]{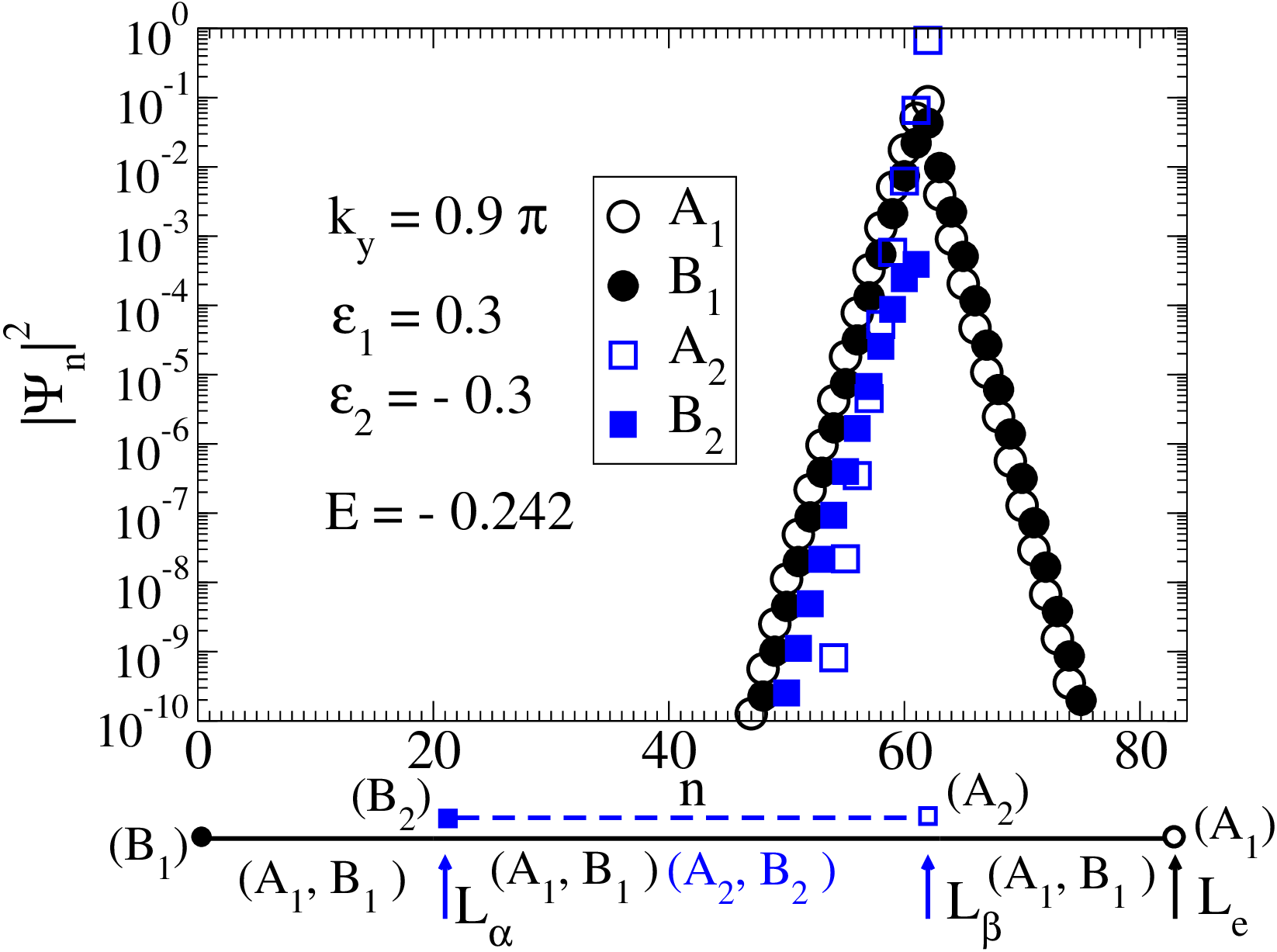}
\end{center}
\flushleft{(c) \hfill (d) \hfill \ }  \\ \vspace{-0.2cm}
\begin{center}
\includegraphics[width=0.23\textwidth]{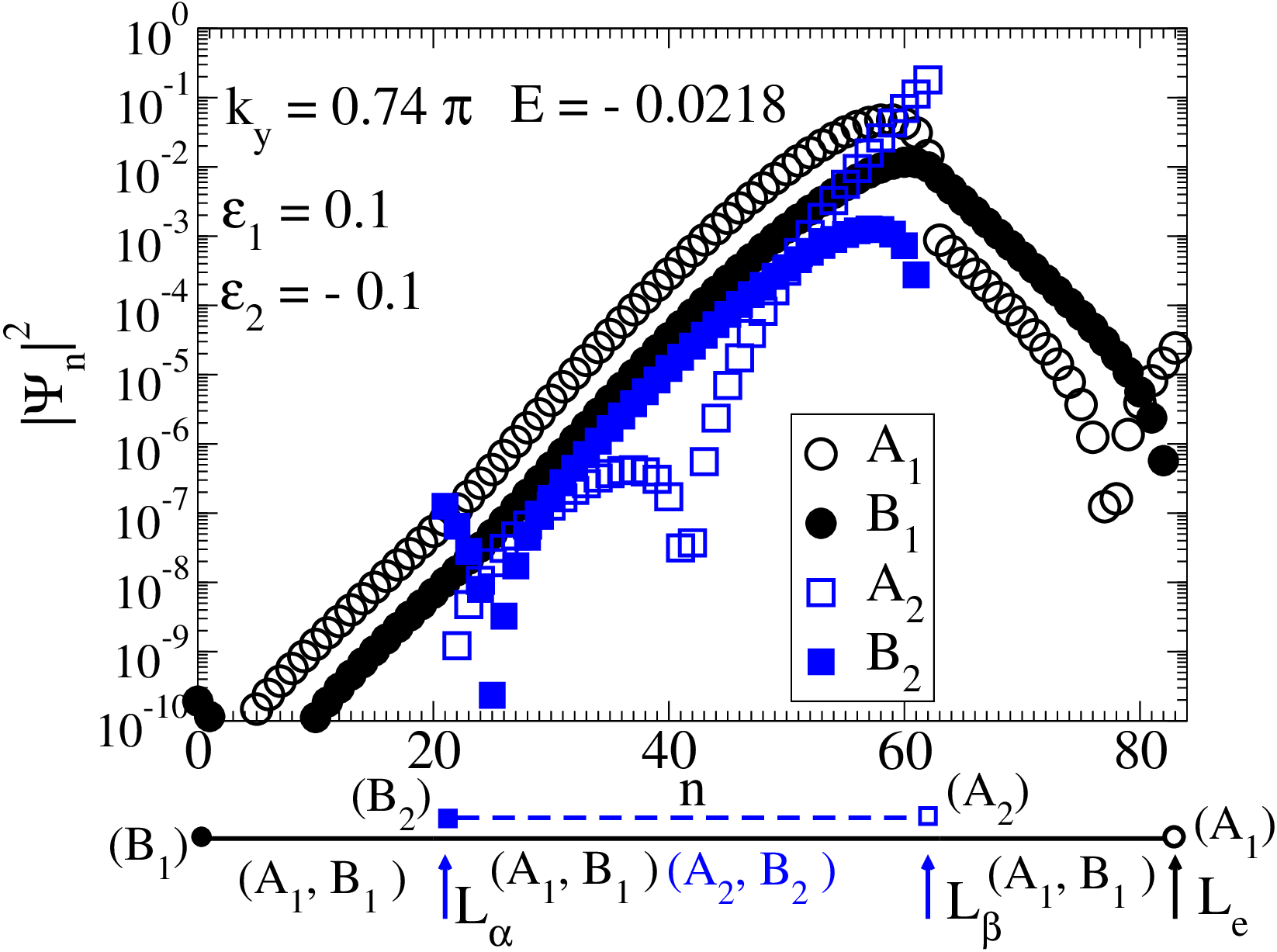} \hfill 
\includegraphics[width=0.23\textwidth]{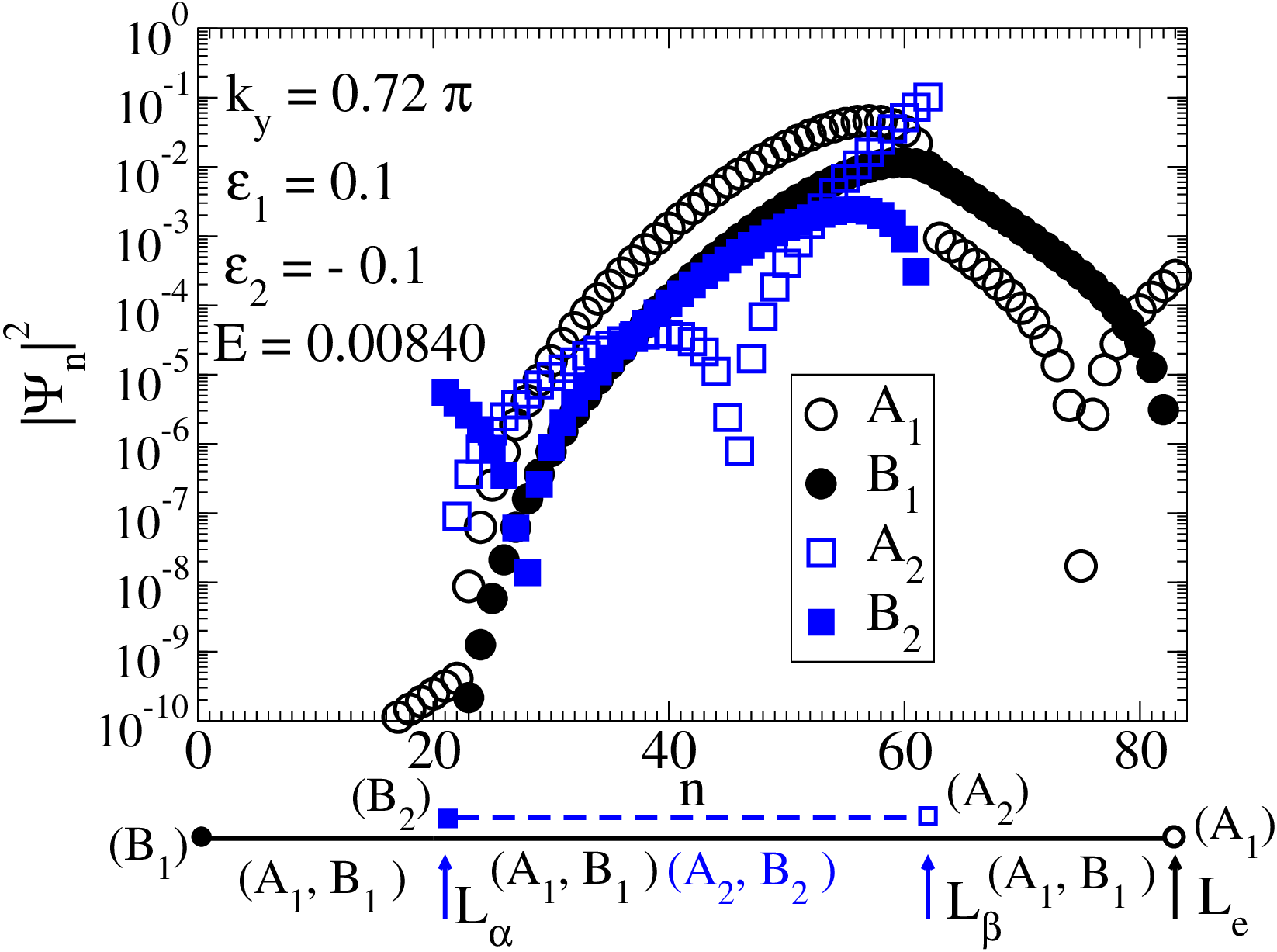}
\end{center}
\caption{(color online).
Edge states at 
the $\beta$-boundary in single-double-single layer graphene at 
$k_y = 0.9 \pi$ with $\epsilon_1=-\epsilon_2=0.1$, (a) and $0.3$ (b).
The localization length of the edge states becomes large 
when $k_y$ approaches to $2 \pi/3$. The edge states
with $\epsilon_1=-\epsilon_2=0.1$
at $k_y = 0.74 \pi$ (c) and $k_y = 0.72 \pi$ (d) are shown.}
\label{figwavesde0013}
\end{figure}
%
\begin{figure}[bt]
\flushleft{(a) \hfill (b) \hfill \ }  \\  
\begin{center}
\includegraphics[width=0.23\textwidth]{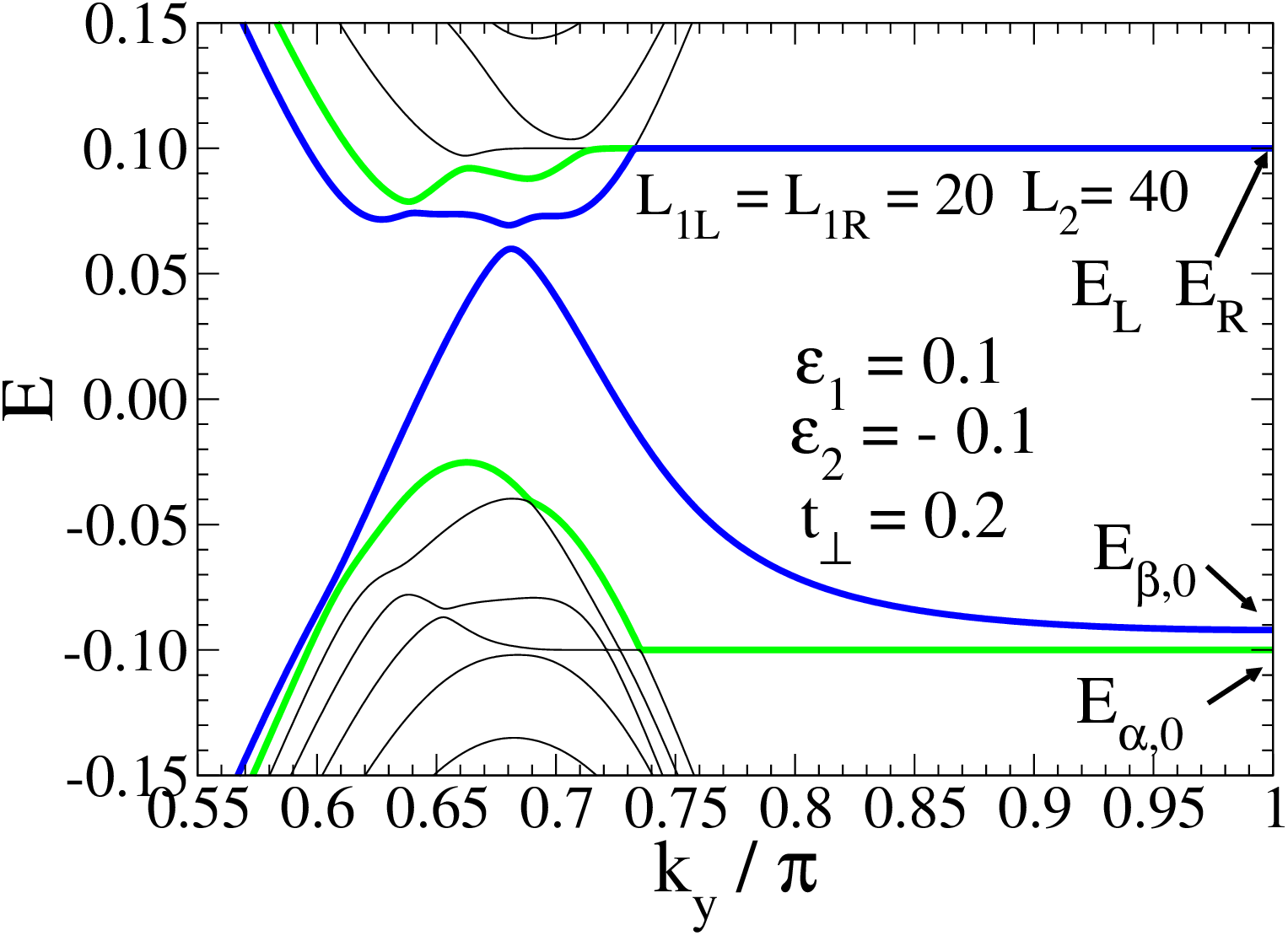} \hfill
\includegraphics[width=0.23\textwidth]{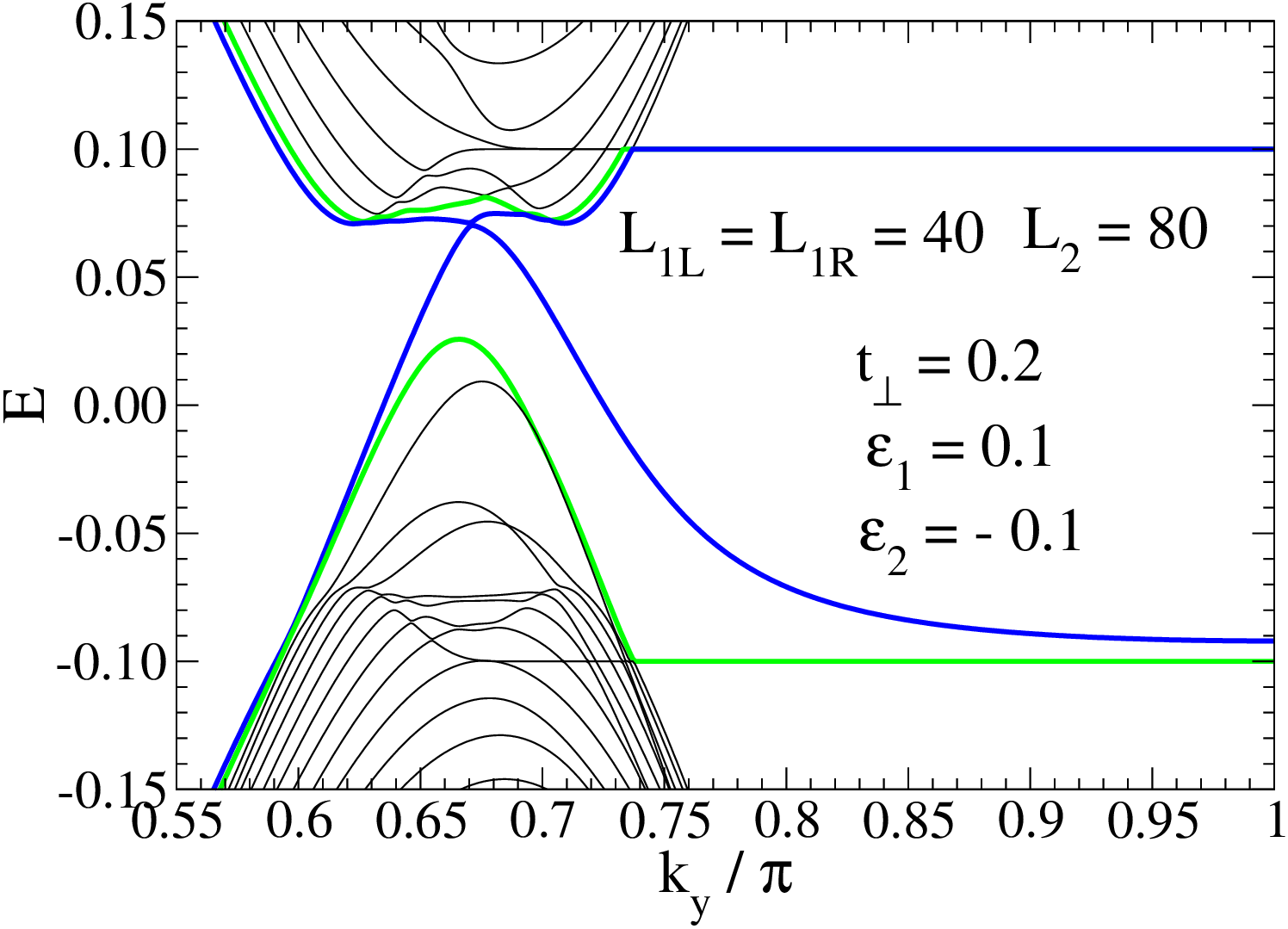} 
\end{center}
\caption{(color online).
Energy of single-double-single layer graphene
as a function of $k_y$ 
with different site energy 
in each layers ($\epsilon_1=-\epsilon_2=0.1 < t_{\perp}$).
Two systems with different width
($L_{1L}=L_{1R}=20$, $L_2=40$ (a) and 
 $L_{1L}=L_{1R}=40$, $L_2=80$ (b) ) are shown. 
The narrower system (a)
is the same but a close up of Fig.~\ref{figsdse01}.
The bulk energy gap (the gap between the upper blue line 
and lower green line at $k_y \approx 2 \pi/3$) becomes smaller
as the width of the system becomes wider.
There exists a small energy gap 
at $k_y \approx 2 \pi /3$ 
between two blue lines in (a), because the localization length of the edge 
state at the $\beta$ edge becomes comparable to the width of the system.
For the wider system (b), 
the energy gap between two blue lines becomes negligible. 
}
\label{figfig6}
\end{figure}

In this way the strictly localized state at the $\beta$-boundary
at $k_y=\pi$ ($a=0$) and $E=E_{\beta,0}$ becomes the localized states
which have the finite amplitudes both in the single-layer
and double-layer regions
when $|2\cos k_y/2| <1-|e_0|$.
In Fig.~\ref{figsdse01} we plot the energy as a function of $k_y$
in the case of the finite 
energy difference between the first layer and the second layer.
The energy of the edge states at the $\beta$-boundary depends on $k_y$
as shown in Fig.~\ref{figsdse01}.
 In Fig.~\ref{figwavesde0013} (a) and (b)  we plot 
the wave functions of the edge states at $E \approx E_{\beta,0}$
and $k_y=0.9 \pi$ for $\epsilon_1=-\epsilon_2=0.1$ and $0.3$.
In contract to the case of 
$\epsilon_1=\epsilon_2=0$ (Fig.~\ref{figwavesdse000} (c)),
all components of the wave functions are finite 
in the double-layer region 
at the $\beta$-boundary ($ n \leq L_{\beta}$).
The wave function has finite amplitudes also in the single-layer region
($ n \geq L_{\beta}+1$),
which means the penetration of the edge state at the $\beta$-boundary
is induced by the electric field.
When $k_y$ approaches to $2 \pi/3$, the energy of the edge states
(the lower blue line)
deviates from $E_{\beta,0}$, as seen in Fig.~\ref{figsdse01}.
The localization length of the edge states becomes large as 
 $k_y$ approaches to $2 \pi/3$ 
(see Fig.~\ref{figwavesde0013} (c) and (d)).
The effects of the $\alpha$-boundary ($n=L_{\alpha}$) and the
right edge ($n=L_e$) on the edge state at $\beta$-boundary
are seen in Fig.~\ref{figwavesde0013} (c) and (d).  

We mention the bulk energy gap caused by the perpendicular electric field.
Though the bulk energy gap is
originated in the double-layer region,
the system with finite width of the single-layer regions
in both side as shown in Fig.\ref{figlatbising2} has the bulk energy gap,
since the the states except for the edge states are extended
in both single-layer and double-layer regions.
When the widths of the single-layer regions 
become larger, the bulk energy gap becomes smaller,
as shown in Fig.~\ref{figfig6} (a) and (b),
which show the $k_y$-dependences of the energies
for the systems with different width.
Therefore, the critical value of $k_y$,
at which the edge state at 
the $\beta$-boundary has the energy between the bulk energy gap,
depends on the width of the system.
The edge states at $k_y= 0.74 \pi$ and $0.72\pi$ 
(Fig.~\ref{figwavesde0013} (c) and (d)) 
have energies between bulk gap ($-0.0252 < E < 0.693$)
for the system with
$L_{1L}=L_{1R}=20$ and $L_{2}=40$, while the edge states at $k_y=0.9 \pi$
(Fig.~\ref{figwavesde0013} (a)) has the energy outside of the bulk gap.
Since STS probes the local density of states\cite{Ritter2009,Tao2011},
the edge states can be observed even if they are outside of the bulk gap. 
Therefore, the field-induced penetration of the edge state into the 
single-layer region can be observed by STS.

As shown in Fig.~\ref{figsdse01} (a),
Fig.~\ref{figfig6} (a) and (b), the edge states  
at the $\beta$-boundary have energies in the bulk energy gap
at $|k_y| \gtrsim 2\pi/3$ if $|\epsilon_1-\epsilon_2| \lesssim 2 t_{\perp}$.
In this case, 
the edge states are partially filled
in the less than half-filled systems. Then electric
 current flow along the $\beta$-boundary ($y$ direction)
in both single-layer region and double-layer region
(see Fig.~\ref{figwavesde0013} (a), (c), and (d)).
The edge states are partially filled even when 
the perpendicular electric field is not applied.
However, the contribution of the edge states 
at the $\beta$-boundary to the electric current 
is only in the double-layer region 
in the absence of the perpendicular electric field,
since the edge states are localized only in the double-layer
region of the $\beta$-boundary 
(see Fig.~\ref{figwavesdse000} (c)). 
Thus, the electric conductivity
along the $\beta$-boundary in the single-layer region
at the $\beta$-boundary
is increased by 
the perpendicular electric field.

\section{conclusion}
We have studied the edge states in the hybrid system of the single-layer
and double-layer graphene. 
By using the tight-binding model, we obtain the analytic solution of the
edge states at the boundary between single-layer and double-layer regions
when perpendicular electric field is not applied.
We found that the edge states at the $\alpha$-boundary are localized 
only in the second layer, and the edge states at the $\beta$-boundary
have finite amplitude in both layers in the double-layer region.

We also find the new edge states with $E \approx E_{\beta,\pm}$, 
which are localized at the $\beta$-boundary.

When the perpendicular electric field is applied, 
the edge states at the $\beta$-boundary are shown to change
drastically. The edge states at the $\beta$-boundary have finite
amplitudes at all sites in both regions of the boundary.
The penetration of the edge states induced by the 
electric field can be observed experimentally 
by STS\cite{Ritter2009,Tao2011}.
The bulk gap becomes smaller for the systems with the 
wider width of the single-layer regions. The edge states,
however, are possible to 
be observed by STS even though the energy of the edge states
are not located in the middle of the bulk gap
because the edge states has larger amplitudes of the wave functions
 than that for the extended states
with the same energy in the region near the boundary.

We propose a simple method to observe the electric-field-induced 
penetration of the edge states.
The conductivity between two terminals 
placed at the single-layer region at the 
$\beta$-boundary will
become large when the perpendicular electric field is applied, as
the edge states at the $\beta$-boundary 
penetrate into the single-layer region.
This electric-field-induced penetration of the edge 
states can be used
as electrical devices.
%

\appendix
\section{analytical solutions for the edge states with $E=0$ at the 
$\alpha$ and $\beta$ boundaries when $\epsilon_1=\epsilon_2$}
\label{appA}

 The equations in the double-layer region ((\ref{eqdouble1})
- (\ref{eqdouble4})) are decoupled into two groups, 
when $\epsilon_{1}=\epsilon_{2}=0$
and $E=0$.
Eqs.~(\ref{eqdouble1}) and (\ref{eqdouble3}) are written as
\begin{equation} 
\left( \begin{array}{cc}
 1 & 0\\
\frac{t_{\perp}}{t} & 1
\end{array} \right)  
\left( \begin{array}{c}
 \Psi_{B_1,n} \\ \Psi_{B_2,n} 
\end{array} \right)
= \left( -2 \cos \frac{k_y}{2} \right)
 \left( \begin{array}{c}
 \Psi_{B_1,n-1} \\ \Psi_{B_2,n-1} 
\end{array} \right) ,
\end{equation}
where $L_{\alpha} +1 \leq n \leq L_{\beta}-1$.
From this equation we obtain
\begin{equation}  
\left( \begin{array}{c}
 \Psi_{B_1,L_{\alpha}+j} \\ \Psi_{B_2,L_{\alpha}+j} 
\end{array} \right)
= \left( -2 \cos \frac{k_y}{2} \right)^j
\left( \begin{array}{cc}
 1 & 0 \\
 -\frac{t_{\perp}}{t} j & 1
\end{array} \right)
 \left( \begin{array}{c}
 \Psi_{B_1,L_{\alpha}} \\ \Psi_{B_2,L_{\alpha}} 
\end{array} \right),
\end{equation}
where $0 \leq j \leq L_2$.
Since we obtain 
\begin{align}
 \Psi_{B_1,L_{\alpha}}
&= \left( -2 \cos \frac{k_y}{2} \right)^{L_{\alpha}} \Psi_{B_1,0},
\end{align}
from the equations in the single-layer region (Eq.~(\ref{eqsingle1})),
 we can take $\Psi_{B_1,L_{\alpha}} = 0$ for $|2 \cos(k_y/2) | <1$ and 
$L_\alpha \gg 1$. 
Therefore,
the edge state localized at the $\alpha$-boundary are obtained as
\begin{equation}
 \left( \begin{array}{c}
  \Psi_{B_1,L_\alpha+j} \\
  \Psi_{B_2,L_\alpha+j} 
\end{array} \right) = 
\left( \begin{array}{c} 
0 \\
\left( -2 \cos \frac{k_y}{2} \right)^j
 \Psi_{B_2,L_\alpha}
\end{array} \right),
\end{equation}
where $0 \leq j \leq L_2$ and other components of $\Psi$ are zero.

In the same way we obtain from
equations (\ref{eqdouble2}) and (\ref{eqdouble4}),
\begin{equation} 
\left( \begin{array}{c}
 \Psi_{A_1,L_{\beta}-j} \\ \Psi_{A_2,L_{\beta}-j} 
\end{array} \right)
= \left( -2 \cos \frac{k_y}{2} \right)^j
\left( \begin{array}{cc}
1 & -\frac{t_{\perp}}{t} j \\
0 & 1
\end{array} \right)
 \left( \begin{array}{c}
 \Psi_{A_1,L_\beta} \\ \Psi_{A_2,L_\beta} 
\end{array} \right),
\end{equation}
where $0 \leq j \leq L_2$.
As in the edge state in the $\alpha$-boundary, we can take
$\Psi_{A_1,L_{\beta}+1}=0$, when  $|2 \cos(k_y/2)| <1$ and $L_{1R} \gg 1$.
Then we obtain from Eq.~(\ref{eqbeta2}) 
\begin{equation}
 \Psi_{A_1,L_{\beta}}=- \frac{t_{\perp}}{t} \Psi_{A_2,L_{\beta}},
\end{equation}
and we obtain the edge state at the $\beta$-boundary as
\begin{equation} 
\left( \begin{array}{c}
 \Psi_{A_1,L_{\beta}-j} \\ \Psi_{A_2,L_{\beta}-j} 
\end{array} \right)
= \left( -2 \cos \frac{k_y}{2} \right)^j \Psi_{A_2,L_\beta}
 \left( \begin{array}{c}
 -\frac{t_{\perp}}{t} (1+j) \\ 1 
\end{array} \right),
\label{eqedgebetaA}
\end{equation}
where $0 \leq j \leq L_2$ and other components of $\Psi$ are zero.

\section{analytical solutions for the edge states with 
$E \approx E_{\beta,\pm}$
at the $\beta$ boundary}
\label{appB}
By replacing $n$ by $n+1$ in Eq.~(\ref{eqsingle1}), we can write
the equations in the single-layer regions,
Eqs.~(\ref{eqsingle1}) and (\ref{eqsingle2}), as
\begin{equation}
 \left( \begin{array}{cc}
 e_0 & 1 \\
 a   & 0 
 \end{array} \right)
\left( \begin{array}{c}
 \Psi_{A_1, n+1} \\
 \Psi_{B_1, n+1}
 \end{array} \right) =
 \left( \begin{array}{cc}
 0 & a \\
 1 & e_0 
 \end{array} \right)
\left( \begin{array}{c}
 \Psi_{A_1, n} \\
 \Psi_{B_1, n}
 \end{array} \right),
\label{eqsingleM}
\end{equation}
where
\begin{align}
 a &= -2 \cos \frac{k_y}{2} ,\\
 e_0 &= \frac{E-\epsilon_1}{t} .
\end{align}
In Eq.~(\ref{eqsingleM}), we can take 
$0 \leq n \leq L_{\alpha}$ or 
$L_{\beta} +1 \leq n \leq L_e -1$, and $\Psi_{A_1,0}=\Psi_{B_1,L_e}=0$.
Then we obtain 
\begin{equation}
\left( \begin{array}{c}
 \Psi_{A_1, n+1} \\
 \Psi_{B_1, n+1}
 \end{array} \right) =
  T 
\left( \begin{array}{c}
 \Psi_{A_1, n} \\
 \Psi_{B_1, n}
 \end{array} \right),
\label{eqrecursive1}
\end{equation}
where $T$ is a $2 \times 2$ matrix given by
\begin{equation}
T=
 \left( \begin{array}{cc}
  \frac{1}{a}   & \frac{e_0}{a} \\
 -\frac{e_0}{a} & \frac{a^2-e_0^2}{a}
 \end{array} \right).
\label{eqmatrixT}
\end{equation}
If $e_0 \neq 0$, the matrix $T$ is diagonalized
by the matrix $V$ as
\begin{equation}
 V^{-1} T V = \left(
\begin{array}{cc}
  \lambda_{+} & 0           \\
   0          & \lambda_{-}
 \end{array} \right),
\label{eqdiag1}
\end{equation}
where 
$\lambda_{\pm}$ are
the eigenvalues of the matrix $T$, 
\begin{equation}
 \lambda_{\pm}= \frac{1}{2a} \left( a^2-e_0^2+1 \pm \sqrt{D} \right),
\label{eqlambda}
\end{equation}
\begin{equation}
 V=\left( \begin{array}{cc}
  \sqrt{D}-a^2+e_0^2+1 & -2e_0 \\
 -2e_0                 & \sqrt{D}-a^2+e_0^2+1 
\end{array} \right),
\label{eqV}
\end{equation}
\begin{equation}
 V^{-1}= \frac{1}{C} \left( \begin{array}{cc}
  \sqrt{D}-a^2+e_0^2+1 & 2e_0 \\
 2e_0                  & \sqrt{D}-a^2+e_0^2+1 
\end{array} \right),
\end{equation}
\begin{equation}
 C=2 \sqrt{D}\left( \sqrt{D} -a^2 + e_0^2 + 1 \right),
\end{equation}
and
\begin{equation}
 D= (a+e_0+1)(a-e_0+1)(a+e_0-1)(a-e_0-1).
\end{equation}
Note that
\begin{equation}
 \lambda_{+} \lambda_{-} =1.
\end{equation}
If $D < 0$, $\lambda_{-}=\lambda_{+}^*$ 
and $|\lambda_{+}|=|\lambda_{-}|=1$.
In this case we obtain the extended states, 
if the boundary conditions at $n=0$, $L_{\alpha}$, 
$L_{\beta}$, and $L_e$ are satisfied.
On the other hand, if $D > 0$, two eigenvalues of $T$ are real 
and either $|\lambda_{+}|$ or $|\lambda_{-}|$ is smaller than 1.
In this case $|\Psi_{A_1,L_\beta+j}|$ and $|\Psi_{B_1,L_\beta+j}|$
can decrease as $|\lambda_+|^j$ $|\lambda_-|^j$ 
obtained from Eq.~\ref{eqrecursive1}, if the boundary 
conditions at $n=L_{\beta}$ are satisfied by the 
corresponding eigenstate of $T$.
Note that $D>0$ is obtained if and only if
\begin{equation}
 | |a| -|e_0| | > 1.
\label{eqeqB13}
\end{equation} 

Now we examine the edge states at the 
$\beta$-boundary with $E \approx
E_{\beta,\pm}$. 
When $k_y \approx \pi$ and 
$E \approx E_{\beta,\pm} = \pm \sqrt{t^2+t_{\perp}^2}$,
we obtain
\begin{equation}
 | a | \ll 1
\end{equation}
and
\begin{equation}
 |e_0| \approx \sqrt{1+\left(\frac{t_\perp}{t}\right)^2} >1.
\end{equation}
Then the inequality Eq.~(\ref{eqeqB13})
 is satisfied and the edge states can exist.
In order to examine the edge states with $E \approx E_{\beta,\pm}$,
we expand $D$, $\lambda_{\pm}$ and $V$ 
in $a$ when $|a| < |e_0|-1$, as
\begin{equation}
 D \approx (e_0^2-1)^2-2 (e_0^2+1) a^2,
\end{equation}
\begin{equation}
 \lambda_{+} \approx - \frac{a}{e_0^2-1},
\end{equation}
\begin{equation}
 \lambda_{-} \approx - \frac{e_0^2-1}{a},
\end{equation}
and
\begin{equation}
 V \approx
(-2e_0)
\left( \begin{array}{cc}
 -e_0 & 1 \\
 1 & -e_0
\end{array} \right).
\end{equation}
In this case $\lambda_{\pm}$ are real and
\begin{equation}
  |\lambda_{+}| < 1 < |\lambda_{-}|.
\end{equation}
The eigenvector of the matrix $T$ with the eigenvalue $\lambda_+$ is
given by the first column of matrix $V$. Therefore,
when $\Psi_{A,L_\beta+1}$ and $\Psi_{B,L_\beta+1}$
($0 \leq j \leq L_{1R}$) satisfy the equation,
\begin{equation}
 \frac{\Psi_{A_1,L_\beta+1}}{\Psi_{B_1,L_\beta+1}}
= - \frac{ \sqrt{D}-a^2+e_0^2+1}{ 2e_0},
\end{equation}
we obtain from Eq.~\ref{eqrecursive1}
\begin{align}
 \left( \begin{array}{c}
  \Psi_{A_1,L_{\beta}+1+j} \\
  \Psi_{B_1,L_{\beta}+1+j} 
 \end{array} \right)
 &= \lambda_+^{j}  \Psi_{B_1,L_\beta+1}
 \left( \begin{array}{c}
 -\frac{\sqrt{D}-a^2+e_0^2+1}{2e_0} \\ 1
 \end{array} \right)
\nonumber \\
& \approx \lambda_+^{j}  \Psi_{B_1,L_\beta+1}
 \left( \begin{array}{c}
 -e_0 \\ 1
 \end{array} \right),
\end{align}
where $0 \leq j \leq L_{1R}$.
This state is the edge state localized 
at the $\beta$-boundary.

At the $\beta$-boundary 
the equations for $\Psi_{A_1,L_{\beta}}$, $\Psi_{B_1,L_{\beta}}$,
$\Psi_{A_2,L_{\beta}}$, $\Psi_{A_1,L_{\beta}+1}$, 
and $\Psi_{B_1,L_{\beta}+1}$ are obtained from  Eq.~(\ref{eqsingle1})
with $n=L_\beta+1$ and Eq.~(\ref{eqbeta2}),
\begin{equation}
 \left( \begin{array}{cc}
 e_0 & 1 \\
 a & 0
 \end{array} \right) 
\left( \begin{array}{c}
\Psi_{A_1,L_{\beta}+1} \\ \Psi_{B_1,L_{\beta}+1}
\end{array} \right)
 = 
 \left( \begin{array}{ccc}
  0 & a   & 0 \\
  1 & e_0 & \frac{t_{\perp}}{t} 
 \end{array} \right) 
  \left( \begin{array}{cc}
\Psi_{A_1,L_{\beta}} \\ \Psi_{B_1,L_{\beta}} \\ \Psi_{A_2,L_{\beta}} 
\end{array} \right).
\end{equation}
This equation is written as
\begin{equation}
\left( \begin{array}{c}
\Psi_{A_1,L_{\beta}+1} \\ \Psi_{B_1,L_{\beta}+1}
\end{array} \right)
 = 
 \left( \begin{array}{ccc}
  \frac{1}{a}    & \frac{e_0}{a} & \frac{t_{\perp}}{at} \\
  -\frac{e_0}{a} & \frac{a^2-e_0^2}{a} & - \frac{e_0 t_{\perp}}{at} 
 \end{array} \right) 
  \left( \begin{array}{cc}
\Psi_{A_1,L_{\beta}} \\ \Psi_{B_1,L_{\beta}} \\ \Psi_{A_2,L_{\beta}} 
\end{array} \right).
\label{eqLbeta1}
\end{equation}
This boundary condition 
as well as the condition that 
the wave functions should decrease exponentially in the
double-layer region in the left part 
(the double-layer region) of the $\beta$-boundary
can be satisfied by adjusting 
$e_0$, $\Psi_{A_1,L_\beta}$, $\Psi_{B_1,L_\beta}$,
and $\Psi_{A_2,L_\beta}$.
It is indeed possible as seen in Fig.~\ref{figwavesdse000}(d).
These edge states at $E \approx E_{\beta,\pm}$ are localized 
at the $\beta$-boundary and the wave functions have the finite
amplitudes in both sides (both single-layer and double-layer 
regions) of the $\beta$-boundary as shown in Fig.~\ref{figwavesdse000}(d).

The above method can be applied to the edge states 
at the single-layer region.
We consider the edge states at the left boundary
in the left single region ($n=0$) as an example.
The boundary condition at $n=0$ is given by
\begin{equation}
 \left( \begin{array}{c}
 \Psi_{A_1,0} \\ 
 \Psi_{B_1,0}
 \end{array} \right) 
= \left( \begin{array}{c}
 0 \\
 \Psi_{B_1,0}
 \end{array} \right). 
\end{equation}
This state can be the eigenvector of $T$ only when $e_0=0$ and the 
eigenvalue of this eigenvector is $a$. Therefore, the edge states at the
left boundary in the single-layer region exist only when $E=\epsilon_1$
and $|a|=|2 \cos \frac{k_y}{2}| <1$,
as obtained in the previous section.
The edge states at the right boundary can be similarly studied
by considering the inverse matrix of $T$.

%
\bibliography{yh}

\end{document}